\documentclass[12pt]{report}
\usepackage{epsfig}
\def\thebibliography#1{\leftline{\large\bf References}\list
  {[\arabic{enumi}]}{\settowidth\labelwidth{[#1]}\leftmargin\labelwidth
\advance\leftmargin\labelsep
\usecounter{enumi}}
\def\newblock{\hskip .11em plus .33em minus .07em}
\sloppy\clubpenalty4000\widowpenalty4000}

\newcommand{\fract}[2]{{\textstyle\frac{#1}{#2}}} %%%
\newcommand{\lder}{\buildrel{\leftarrow}\over{\nabla}}
\newcommand{\rder}{\buildrel{\rightarrow}\over{\nabla}}
\newcommand{\ldert}{\buildrel{\leftarrow}\over{\partial}}
\newcommand{\rdert}{\buildrel{\rightarrow}\over{\partial}}
\newcommand{\gapprox}{\, \stackrel{\scriptstyle\raisebox{-3.0mm}{$>$}}
{\scriptstyle \raisebox{-1.5mm}{$\sim$}}\,}
\newcommand{\lapprox}{\, \stackrel{\scriptstyle\raisebox{-3.0mm}{$<$}}
{\scriptstyle \raisebox{-1.5mm}{$\sim$}}\,}

\newcommand{\arsinh}{\sinh^{-1}}

\begin{document}

\begin{center}
{\large \bf
Calculating Vacuum Energies in Renormalizable Quantum Field
Theories: A New Approach to the Casimir Problem}
\end{center}

\centerline{
N.~Graham$^{\rm a}$,
R.L.~Jaffe$^{\rm b}$, V.~Khemani$^{\rm b}$,
M.~Quandt$^{\rm b}$, M.~Scandurra$^{\rm b}$,
H.~Weigel\footnote{Heisenberg Fellow}$^{\rm b,c}$
}

\parbox[t]{15cm}{
\begin{center}
{~}\\$^{\rm a}$Department of Physics and Astronomy,\\
University of California, Los Angeles \\
Los Angeles, CA  90095 \\~\\
$^{\rm b}$Center for Theoretical Physics,
Laboratory for Nuclear Science\\ and Department of Physics,
Massachusetts Institute of Technology\\ Cambridge, Massachusetts 02139 \\~\\
$^{\rm c}$Institute for Theoretical Physics, T\"ubingen University\\
D-72076 T\"ubingen, Germany\\
{~} \\
{\qquad \rm UCLA/02/TEP/14 \qquad MIT-CTP-3278 \qquad UNITU-HEP-11/2002 \\
hep-th/0207120}
\end{center}
}

\centerline{\large\bf Abstract}

{\small

The Casimir problem is usually posed as the response of a fluctuating
quantum field to externally imposed boundary conditions.  In reality,
however, no interaction is strong enough to enforce a boundary
condition on all frequencies of a fluctuating field.  We construct a
more physical model of the situation by coupling the fluctuating field
to a smooth background potential that implements the boundary
condition in a certain limit.  To study this problem, we develop
general new methods to compute renormalized one--loop quantum
energies and energy densities.  We use analytic properties of
scattering data to compute Green's functions in time--independent
background fields at imaginary momenta.  Our calculational method is
particularly useful for numerical studies of singular limits because
it avoids terms that oscillate or require cancellation of
exponentially growing and decaying factors.  To renormalize, we
identify potentially divergent contributions to the Casimir energy
with low orders in the Born series to the Green's function.  We
subtract these contributions and add back the corresponding Feynman
diagrams, which we combine with counterterms fixed by imposing
standard renormalization conditions on low--order Green's functions.
The resulting Casimir energy and energy density are finite functionals
for smooth background potentials.  In general, however, the Casimir
energy diverges in the boundary condition limit. This divergence is
real and reflects the infinite energy needed to constrain a
fluctuating field on all energy scales; renormalizable quantum field
theories have no place for ad hoc surface counterterms.  We apply our
methods to simple examples to illustrate cases where these subtleties
invalidate the conclusions of the boundary condition approach.
}

\leftline{\it \small Keywords:~\parbox[t]{15cm}{
Energy densities, Green's functions, density of states,
renormalization, Casimir effect}}

\leftline{\it \small PACS:~\parbox[t]{15cm}{
03.65.Nk, 03.70.+k, 11.10.Gh}}

\newpage

\bigskip
\stepcounter{section}
\leftline{\large\bf 1. Introduction}
\bigskip

Long ago, Casimir predicted the existence of a force generated by
quantum fluctuations of the electromagnetic field between uncharged,
perfectly conducting metal plates in vacuo~\cite{Casimir}.  Since
then, many related calculations have been carried out (for reviews
see~\cite{Grib,MT,MiltonReview,Plunien,Leipzig-proceedings}).  The standard
approach to problems of this type has been to consider the response of
a free quantum field to time--independent boundary conditions imposed
on stationary surfaces.  This calculation is an idealization of the
more physical description of the quantum field as interacting with a
smooth external potential, which goes to zero away from the
surfaces. This paper will provide the technical tools necessary for
understanding precisely when this idealization is justified, and when
it is hazardous.  We will demonstrate the use of these tools in some
simple models; the reader is also referred to \cite{letter} for a
briefer, less technical, and more focused discussion of a variety of
Casimir problems.

In the field theory picture, the dynamics of a fluctuating field in a
background potential is governed by the one--loop effective energy, which
can be expressed either as an infinite sum of Feynman diagrams or as a
``Casimir sum'' over modes.  Both expressions are formally divergent, so
they can only be defined by introducing counterterms fixed with physical
renormalization conditions.  A Dirichlet boundary condition emerges when a
repulsive background scalar field becomes strong and sharply peaked. We
will see that the renormalized total energy of the
fluctuating field relative to its value in the absence of boundaries can
diverge in this limit.  In such cases, the idealized Casimir
energy {\it per se\/} does not exist.  However, these divergences are
associated with the boundaries and therefore do not affect the energy
density away from the boundaries or the force between rigid surfaces.
Nevertheless some geometric quantities, such as the ``Casimir
stress''~\cite{Boyer} --- the force per unit area on the surface where
the boundary conditions are applied --- are sensitive to these
divergences.  In such cases, there is no way to calculate meaningful
finite results by considering boundary conditions.

To demonstrate these results in detail, we require efficient and robust
techniques for computing Casimir energies and energy densities, which
we describe in Section 2.  The calculation of Casimir energy
densities is interesting in its own right, for example in the
study of energy conditions in general
relativity \cite{Grib,Birrel,NoahKen}.  Our approach will phrase the
calculation in the language of conventional renormalizable quantum field
theory, in a way that is amenable to efficient numerical computation.
We study the vacuum polarization energy and energy density of a
fluctuating boson field $\phi$ coupled to a background $\sigma(\vec
x)$, which is sharply peaked in the boundary condition limit.
We introduce counterterms fixed by perturbative renormalization
conditions, which define physical inputs to the theory such as
particle masses and coupling constants at particular values of the
external momenta.  The resulting renormalization scheme is
conventional, precise, and unambiguous.   It can be related to any
another conventional scheme by the analysis of low--order Feynman
diagrams.  We do not, however, see any relation between this scheme and
analyses in which boundary surface dependent counterterms are
introduced in an {\it ad hoc\/} manner.  It is a common
procedure in the literature to simply ignore divergent surface
integrals and pole terms arising in the course of the calculation
\cite{Mamayev, Milton}.  The $\zeta$--function method, in particular,
introduces counterterms of any power of the geometrical parameter
\cite{Blau}.  Such counterterms redefine classical parameters of
the background, but have no justification in quantum field theory.  As
a result, these approaches do not appear to be reconcilable with ordinary
renormalizable quantum field theory.

Our method is based on an extension of methods developed for the total energy
\cite{method,method1}, together with an extension of the analytic
continuation methods introduced in Ref.~\cite{BK}.  First, we relate
the matrix element of the energy density operator to the
scattering Green's function at coincident points, $G (\vec x,\vec
x,k)$.  We regulate the large $k$ behavior of $G$ by subtracting the
first few terms in its Born expansion and adding back the contribution
of the associated low--order Feynman diagrams \cite{method,method1}.
The Green's function is difficult to deal with at large (real) $k$
when the background field approaches the boundary condition
limit: it oscillates with amplitude much larger than the net
contribution to the energy density.  To avoid such oscillations
we rotate $k$ to the imaginary axis \cite{BK, BL}.  Before doing so,
however, we rewrite the Green's function as a product of
terms, each of which is bounded for $k$ on the imaginary axis and
each of which can be easily computed by integrating a
Schr\"odinger--like equation subject to simple boundary
conditions.  This formalism allows us to avoid
exponentially growing quantities anywhere in the calculation.
As a result, we are able to write the Casimir energy density as an
integral along the positive imaginary axis and perform the integral
directly for imaginary $k$.  Here we depart from Refs. \cite{BK,
BL}, which formulate the problem as an integral along the imaginary axis
but then use a dispersion relation to compute the integrand in terms
of scattering data for real $k$.

The final ingredient we need for the calculation is the Feynman
diagram contribution, which is where we implement the renormalization
conditions.  The energy density operator $\hat{T}_{00}$ generates
insertions in Feynman diagrams of every order in the background field
$\sigma$.  Using a functional approach allows us to keep careful track
of orders in $\sigma$ in this calculation, which we need to maintain
consistency with the rest of the calculation.

In Section 3 we show how to apply our methods to a simple example, the
Casimir energy density for a boson field in one space dimension.  We
couple the fluctuating field $\phi$ to a background $\sigma(x)$ with
an interaction of the form $\lambda\phi^{2}\sigma(x)$.  For $\sigma(x)
= \delta(x-a)+\delta(x+a)$, we obtain Dirichlet boundary conditions at
$x=\pm a$ in the limit where the coupling constant $\lambda$ becomes
infinite.  We find that the \emph{total} Casimir energy diverges in
this limit.  The divergence is localized at the points where the
background $\sigma$ is nonzero.  For all $x\ne\pm a$, the energy
density remains finite and as $\lambda$ approaches infinity it
smoothly approaches the result obtained by imposing the boundary condition
$\phi=0$ at $x=\pm a$.  To complete Section 3 we study the vacuum
polarization energy in the presence of a smooth, strongly peaked
background field.  By taking $\sigma$ to be the sum of Gau{\ss}ians
centered at $\pm a$, we can study the way that the energy density
behaves as $\sigma(x)$ approaches the boundary condition limit.  As
expected, it approaches its limiting form at every $x\ne\pm a$, though
the rate of approach depends on the proximity to $\pm a$.

In Section 4 we consider the case of a circle in two spatial
dimensions.  We consider Gau{\ss}ian backgrounds peaked around $r=a$
and study the limit in which the Gau{\ss}ian approaches a
$\delta$-function.  When the strength of the coupling goes to
infinity, we obtain a Dirichlet boundary condition at $r=a$.  However,
the renormalized total energy diverges in the $\delta$-function limit,
even at finite coupling.  This result is a simple consequence of the
fact that $\int d^{2}x[\sigma(r)]^{2}\to\infty$ in this limit.  As a
result, the surface tension on the circle --- the derivative of the
energy with respect to $a$ --- diverges.  This example illustrates the
generic problem we have found with attempts to define the Casimir
energy in the context of a quantum field theory: Even though the
renormalized vacuum polarization energy is a finite functional of a
smooth background, it diverges as the background field assumes the
singular configuration necessary to implement the boundary condition. 
This divergence affects any physical quantity, such as the surface
tension, whose measurement requires comparison of configurations with
different boundaries, for which the divergent contributions are
different.  We take this divergence as an indication that such
quantities depend on the details of the interaction between the
fluctuating field and the material to which it couples.  Thus
the Casimir stress cannot be defined in a way that is independent of
the other stresses to which the material is subject.

We close with a summary and some considerations for future work in
Section 5 and give technical details in the Appendices.

\bigskip

\stepcounter{section}
\leftline{\large\bf 2.  Method}
\bigskip
We begin by developing the method for computing the energy density
of a fluctuating quantum field coupled to a classical background.
From the corresponding scattering and bound state
wavefunctions, we construct a Fock decomposition of the quantum
field and compute the matrix element of the energy density
operator $\hat{T}_{00}$.  We express this matrix element in terms of a
Green's function with appropriate boundary conditions.  The energy
density is given by a sum over bound states plus an integral over
the continuum labeled by the wave number $k$.  We develop a
representation of the Green's function suitable for analytic
continuation into the upper half $k$--plane, so the
$k$ integral can be deformed along a cut on the positive imaginary
axis. To regulate the ultraviolet divergences of the theory, which
corresponds to eliminating the contribution associated with the
semi--circle at infinite complex momenta, we subtract leading Born
approximations to the Green's function, which we later add back in as
Feynman diagrams.  These diagrams are then regularized and
renormalized in ordinary Feynman perturbation theory.

\bigskip

\stepcounter{subsection} \leftline{\bf 2.1 Formalism}
\bigskip

We consider a static, spherically symmetric background potential
$\sigma=\sigma(r)$ with $r=|\vec{x}|$ in $n$ spatial dimensions.
The symmetric energy density operator for a real scalar field coupled
to $\sigma$ is 
\begin{eqnarray}
\hat{T}_{00}(x)&=&\frac{1}{2}\left[
\dot\phi^2+(\vec{\nabla}\phi)^2+m^2\phi^2+\sigma(r)\phi^2\right] \cr
&=&\frac{1}{2}\left[
\dot\phi^2+\phi\left(-\vec{\nabla}^2+m^2+\sigma(r)\right)\phi\right]
+\frac{1}{4}\vec{\nabla}^2\left(\phi^{2}\right)
\label{H_dense}
\end{eqnarray}
where we have rearranged the spatial derivative term in order to be able to
use the Schr\"o\-din\-ger equation to evaluate the expression in
brackets.  We assume that the background potential is smooth, 
with $\int_0^\infty dr |{{\sigma}}(r)| (1+r) <\infty$.\footnote{In
view of later applications with singular background fields, we can
relax this condition to allow for all $\sigma(r)$ for which a
scattering problem in the usual sense can be defined.  However, some
of the standard bounds on the asymptotics of Jost functions no longer
hold for a non--smooth $\sigma(r)$, giving rise to the additional
singularities discussed later.}  We define the ``vacuum'' to be the
state $|\Omega\rangle$ of lowest energy in the background $\sigma$ and
the ``trivial vacuum'' to be the state $|0\rangle$ of lowest energy
when $\sigma\equiv0$. The vacuum energy density is the renormalized
expectation value of $\hat{T}_{00}$ with respect to the vacuum
$|\Omega\rangle$, $\langle\Omega|\hat{T}_{00}(x)|\Omega\rangle_{\rm
ren}$, which includes the matrix elements of the counterterms.  Since
we have spherical symmetry, we will define the energy density in a spherical
shell,
\begin{equation}
\epsilon(r) = \frac{2\pi^{n/2}}{\Gamma(\frac{n}{2})} r^{n-1}
\langle\Omega|\hat{T}_{00}(x)|\Omega\rangle_{\rm ren} \,.
\end{equation}
We first perform a partial wave decomposition using spherical
symmetry,
\begin{equation}
\phi(t,\vec{x})=\sum_{\{\ell\}}\phi_\ell(t,r) {\cal Y}_{\{\ell\}}(\hat{x})
\label{partial_wave}
\end{equation}
where ${\cal Y}_{\{\ell\}}(\hat{x})$ are the $n$--dimensional
spherical harmonics$\{\ell\}$ refers to the set of all
angular quantum numbers in $n$ dimensions.  The total angular
momentum assumes integer values $\ell = 0,1,2,\ldots$ in all
dimensions except for $n=1$, where $\ell=0$ and $1$ only, corresponding
to the symmetric and antisymmetric channels respectively.
We make the Fock decomposition
\begin{eqnarray}
\phi_\ell(t,r)&=& \frac{1}{r^{\frac{n-1}{2}}}
\int_0^\infty \frac{dk}{\sqrt{\pi\omega}}
\left[\psi_\ell(k,r)\,e^{-i\omega t}a_{\ell}(k) +
\psi_\ell^*(k,r)\,e^{i\omega t}a^\dagger_{\ell}(k)\right] \cr
&& + \frac{1}{r^{\frac{n-1}{2}}}\sum_j\frac{1}{\sqrt{2\omega_j}}
\left[\psi_{\ell j}(r)\,e^{-i\omega_j t}a_{\ell j}
+\psi_{\ell j}(r)\,e^{i\omega_j t}a^\dagger_{\ell j}\right]\, ,
\label{psi_decomp}
\end{eqnarray}
which we have split into scattering states with
$\omega=\sqrt{k^2+m^2}$ and bound states with
$\omega_j=\sqrt{m^2-\kappa_j^2}$.  Note that the latter wave functions
are real.  The vacuum $|\Omega\rangle$ is annihilated by all of the
$a_{\ell}(k)$ and $a_{\ell j}$.  The radial wavefunctions in
eq.~(\ref{psi_decomp}) are solutions to the Schr\"odinger--like
equation
\begin{equation} -\psi'' + \frac{1}{r^2}\left(\nu-\frac{1}{2}\right)
\left(\nu+\frac{1}{2}\right)\,\psi +{{\sigma}}(r)\psi - k^{2}\psi =0
\label{Schroedinger}
\end{equation}
where
\begin{equation}
\nu=\ell-1+\frac{n}{2} \,.
\label{defnu}
\end{equation}
In each channel, we normalize the wavefunctions to give the
completeness relation
\begin{equation}
\frac{2}{\pi}\int_0^\infty dk\,
\psi_\ell^*(k,r)\psi_\ell(k,r^\prime)+\sum_j
\psi_{\ell j}(r)\psi_{\ell j}(r^\prime) =\delta(r-r^\prime)
\label{complete}
\end{equation}
which implies the orthonormality relations
\begin{equation}
\begin{array}{r@{\,\,\,=\,\,\,}l@{\qquad}l}
\displaystyle\int_0^\infty \psi_{\ell j}(r) \psi_{\ell j'}(r)\,dr 
& \displaystyle\delta_{jj'} & \mbox{for the bound states and} \\[3mm]
\displaystyle\int_0^\infty \psi_\ell^*(k,r) \psi_\ell(k',r)\,dr 
& \displaystyle\frac{\pi}{2}\delta(k-k') & \mbox{for the continuum states.}
\end{array}
\label{orthonorm}
\end{equation}
For ${{\sigma}}\equiv0$, the normalized wavefunctions are
$\psi^{(0)}_\ell(k,r)=\sqrt{\frac{\pi}{2}kr}J_\nu(kr)$.  With the
above conventions and normalizations, the standard equal time
commutation relations for the quantum field $\phi$ yield canonical
commutation relations for the creation and annihilation operators,
$[a_\ell(k),a^\dagger_{\ell^\prime}(k^\prime)] =
\delta(k-k^\prime)\delta_{\ell,\ell^\prime}$ and $[a_{\ell j},
a^{\dagger}_{\ell'j'}]=\delta_{jj'}\delta_{\ell\ell'}$ while all
other commutators vanish.

The vacuum expectation value of the energy density can now be computed
by using eq.~(\ref{psi_decomp}) in eq.~(\ref{H_dense}). We obtain
\begin{eqnarray}
\epsilon(r)&=&
\sum_\ell N_\ell\left[\int_0^\infty \frac{dk}{\pi}\omega
\psi_\ell^*(k,r)\psi_\ell(k,r)+
\sum_j\frac{\omega_j}{2} \psi_{\ell j}(r)^2\right]
\cr && + \frac{1}{4}D_r \sum_\ell N_\ell
\left[\int_0^\infty\frac{dk}{\pi\omega}
\psi_\ell^*(k,r)\psi_\ell(k,r)+
\sum_j\frac{1}{2\omega_j}\psi_{\ell j}(r)^2 \right]
-\epsilon^{(0)}(r)+\epsilon_{\rm CT}(r)\,.
\label{ebare}
\end{eqnarray}
where $N_\ell= \frac{\Gamma(n+\ell-2)}{\Gamma(n-1)
\Gamma(\ell+1)}(n+2\ell-2)$ is the degeneracy factor in the
$\ell^{\rm th}$ partial wave, $D_r=\frac{\partial}{\partial
r}\left(\frac{\partial}{\partial r} -\frac{n-1}{r}\right)$,
$\epsilon_{\rm CT}(r)$ is the counterterm contribution, and
$\epsilon^{(0)}(r)$ indicates the subtraction of the energy in the
trivial vacuum, which is just given by evaluating the first integral
with free wavefunctions $\psi^{(0)}_\ell(k, r)$.  This subtraction
corresponds to a renormalization of the cosmological constant.

We can identify the scattering state contribution with the Green's function
by defining the local spectral density
\begin{equation}
\rho_\ell(k,r) \equiv \frac{k}{i} G_\ell(r,r,k) \, ,
\label{scatdense1}
\end{equation}
in the upper half--plane, so that for real $k$
\begin{equation}
\mathsf{Re} \left\{\rho_\ell(k,r)\right\} = \psi_\ell^*(k,r)\psi_\ell(k,r) = 
\mathsf{Im} \left\{k\,G_\ell(r,r,k)\right\} \, ,
\label{scatdense2}
\end{equation}
where
\begin{equation}
G_\ell(r,r^\prime,k)=-\frac{2}{\pi}\int_0^\infty dq
\frac{\psi_\ell^*(q,r)\psi_\ell(q,r^\prime)}
{(k+i\epsilon)^2-q^2}
-\sum_j\frac{\psi_{\ell j}(r)\psi_{\ell j}(r^\prime)}
{k^2+\kappa_j^2}\,.
\label{Green1}
\end{equation}
The $i\epsilon$ prescription has been chosen so that this Green's function
is meromorphic in the upper half--plane, with simple poles at the
imaginary momenta $k=i\kappa_j$ corresponding to bound states.  Thus we have
\begin{eqnarray}
\epsilon(r)&=&
\sum_\ell N_\ell \int_{-\infty}^\infty \frac{dk}{2\pi i} \omega
\left[1+ \frac{1}{4\omega^2}D_r\right]
k G_\ell(r,r,k) \nonumber \\*  &&
+\sum_\ell N_\ell \sum_j \omega_j
\left[1+\frac{1}{4\omega_j^2} D_r\right]
\psi_{\ell j}(r)^2 -\epsilon^{(0)}(r)+\epsilon_{\rm CT}(r)\, ,
\label{ebare1}
\end{eqnarray}
where we have made use of the fact that for real $k$, the imaginary
part of the Green's function at $r=r^\prime$ is an odd function
of $k$, while the real part is even. We would like to use
eq.~(\ref{ebare1}) to compute $\epsilon(r)$ as a contour integral in
the upper half--plane.  First we must eliminate the contribution from
the semi--circular contour at large $|k|$ with $\mathsf{Im}(k)\ge0$.
Using the techniques of Refs.~\cite{past1} and~\cite{past2} we see that
subtracting sufficiently many terms in the Born series from the Green's
function yields a convergent integral.  We then add back exactly what
we subtracted in the form of Feynman diagrams.  We define
\begin{eqnarray}
\left[\rho_\ell(k,r)\right]_{N} &\equiv&
\left[\rho_\ell(k,r)-\rho_\ell^{(0)}(k,r)-\rho_\ell^{(1)}(k,r)
\ldots -\rho_\ell^{(N)}(k,r) \right] \cr
&=& \frac{k}{i} \left[G_\ell(r,r,k)-G^{(0)}_\ell(r,r,k)
- G^{(1)}_\ell(r,r,k)-\ldots-G^{(N)}_\ell(r,r,k) \right]
\label{state_dens}
\end{eqnarray}
where the superscript $(j)$ indicates the term of order $j$ in the
Born expansion.  Subtracting the free Green's function
$G^{(0)}_\ell(r,r,k)$ corresponds to subtracting $\epsilon^{(0)}(r)$
above.  We substitute $\left[\rho_\ell(k,r)\right]_{N}$ for
$kG_\ell(r,r,k)/i$ in eq.~(\ref{ebare1}) and add back in the
Feynman diagrams corresponding to the subtractions in
eq.~(\ref{state_dens}), which we compute below.  As a result, we
obtain a finite expression amenable to contour
integration. Furthermore, we have precisely identified the potentially
divergent pieces with Feynman diagrams, which we can regularize and
renormalize using standard methods.  When combined with the
contribution from the counterterms $\epsilon_{\rm CT}(r)$ they yield
finite contributions to the energy density (for smooth backgrounds).

\begin{figure}
 \centerline{
  {\epsfig{file=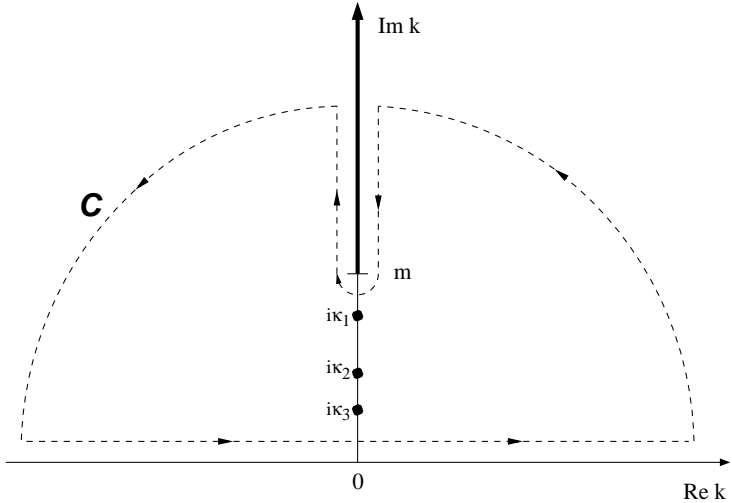,height=7cm,width=10cm}}}
 \caption{\small Contour for $k$ integration.}
\label{contourfig}
\end{figure}

We now have to specify the contour along which to compute the
integral
\begin{equation}
\oint_{\cal C} \frac{dk}{2\pi} \sqrt{k^2+m^2}\left[1+
\frac{1}{4(k^2+m^2)}D_r\right] \left[\rho_\ell(k,r)\right]_{N} \,.
\label{contour}
\end{equation}
% %
The integrand has a branch cut along the imaginary axis $k\in
[im,+i\infty]$ and simple poles at the bound state momenta
$k=i\kappa_j$.  We deform the integral around the cut as shown
in Fig.~\ref{contourfig},
picking up the residues, which cancel the bound state contributions in
eq.~(\ref{ebare1}) \cite{Bordag}.  Note that the Born terms do not
introduce any poles.  Hence we are left with the discontinuity along the cut,
\begin{eqnarray}
\epsilon(r) &=& -\sum_\ell N_\ell \int_m^\infty\frac{dt}{\pi}
\sqrt{t^2-m^2} \left[1-\frac{1}{4(t^2-m^2)}D_r\right]
\left[\rho_\ell(it,r)\right]_{N}
+ \sum_{i=1}^{N} \epsilon_{\rm FD}^{(i)}(r) + \epsilon_{\rm CT}(r) \cr
&\equiv& \bar\epsilon(r) + \sum_{i=1}^{N} \epsilon_{\rm FD}^{(i)}(r) +
\epsilon_{\rm CT}(r) \, .
\label{fundamental}
\end{eqnarray}
Here $\epsilon_{\rm FD}^{(i)}(r)$ is the Feynman diagram contribution
at order $i$ in the background field.  As emphasized in the
Introduction, the representation of  $\epsilon(r)$ as an integral
along the imaginary axis is only useful if it can be computed
efficiently.  In the following subsections we show how to do so.

\bigskip

\stepcounter{subsection}
\leftline{\bf 2.2 Computational Techniques}
\bigskip

In this subsection we concentrate on
$\bar\epsilon(r)$, the continuum integral in
eq.~(\ref{fundamental}). We develop a representation of the
partial wave Green's function, $G_\ell(r,r^\prime,k)$ and its
Born series, eq.~(\ref{state_dens}), that can be computed
efficiently for imaginary $k$.  By working with pure imaginary
momenta we avoid functions that would oscillate at large $|k|$.
We will make extensive use of the analytic properties of scattering
data~\cite{Ch77}.

We begin by introducing solutions to
eq.~(\ref{Schroedinger}) that obey a variety of different
boundary conditions.  First we define the free Jost solution
\begin{equation}
w_\ell(kr)=(-1)^\nu\sqrt{\frac{\pi}{2}kr}
\left[J_\nu(kr)+iY_\nu(kr)\right]\, , \label{freeJ}
\end{equation}
where $\nu=\ell-1+\fract{n}{2}$. The radial function $w_{\ell}(kr)$ is
a solution to eq.~(\ref{Schroedinger}) with ${{\sigma}}\equiv0$
describing an outgoing spherical wave. Then we define
\begin{itemize}
\item The \emph{Jost solution}, $f_\ell(k,r)$. It behaves like an
outgoing wave at $r\to\infty$, with
\begin{equation}
\lim_{r\to\infty} \frac{f_\ell(k,r)}{w_\ell(kr)} = 1\,.
\label{jost_bc}
\end{equation}
\item The \emph{Jost function}, $F_\ell(k)$.  It is obtained as
the ratio of the interacting and free Jost solutions at
$r=0$,
\begin{equation}
F_\ell(k) = \lim_{r\to 0}\frac{f_\ell(k,r)}{w_\ell(kr)}\,.
\label{Jostfunc}
\end{equation}
\item The \emph{regular solution}, $\phi_\ell(k,r)$. It is defined by a
$k$--independent boundary condition at the origin,
\begin{equation}
\lim_{r\to 0} \frac{\Gamma(\nu+1)}{\sqrt{\pi}}
\left(\frac{r}{2}\right)^{-(\nu+\frac{1}{2})} \phi_\ell(k,r) = 1\,.
\label{bcregular}
\end{equation}
\item The \emph{physical scattering solution}, $\psi_\ell(k,r)$.  It
is also regular at the origin (and thus proportional to
$\phi_\ell(k,r)$), but differently normalized\footnote{Whenever a
fractional power of $k$ appears, we define it to be the limit as $k$
approaches the real axis from above.}
\begin{equation}
\psi_\ell(k,r) = \frac{k^{\nu+\frac{1}{2}}}{F_\ell(k)} \phi_\ell(k,r)\, .
\end{equation}
\end{itemize}

The reason for distinguishing two regular solutions is that
$\phi_\ell$ has a simple boundary condition at $r=0$, while
$\psi_\ell$ has a physical boundary condition at $r\to\infty$,
corresponding to incoming and outgoing spherical waves.
With these definitions, the Green's function has a simple representation
\begin{equation}
G_\ell(r,r',k) = \frac{\phi_\ell(k,r_{<}) f_\ell(k,r_{>})}{F_\ell(k)}
(-k)^{\nu-\frac{1}{2}}\, ,
\label{Green2}
\end{equation}
where $r_>$ ($r_<$) denotes the larger (smaller) of the two arguments
$r$ and $r^\prime$.  The poles of $G$ occur at the zeros of the Jost
function, which are the imaginary bound state momenta. Note
that these are the only poles of eq.~(\ref{Green1}) in the upper
half--plane, and since the two functions in eq.~(\ref{Green1})
and eq.~(\ref{Green2}) obey the same inhomogeneous differential
equation they are indeed identical.

The representation of $G_{\ell}(r,r,k)$ in eq.~(\ref{Green2}) is not
yet suitable for numerical computation on the imaginary axis.
Although $G_{\ell}$ is analytic in the upper half--plane, $f_\ell$ and
$\phi_\ell$ contain pieces that oscillate for real $k$ and
exponentially increase or decrease respectively when $k$ has an
imaginary part.  We are actually only interested in the case
$r=r^\prime$, in which case the product $f_\ell\phi_\ell$ is
well--behaved.  But because of the exponential behavior  of the two
factors, the product cannot be computed accurately by simply
multiplying the individual functions obtained from numerical
integration of the respective differential equations.  To proceed, we
factor out the dangerous exponential components with the following
{\it ansatz},\footnote{For $n=1$ and $n=2$, the case of $\ell=0$ is
somewhat different.  We will explore the former in Section~3 and the
latter in Section~4.}
\begin{eqnarray}
f_\ell(k,r) &\equiv& w_\ell(kr)  g_\ell(k,r) \cr
\noalign{ \hbox{and}} \phi_\ell(k,r) &\equiv
&\frac{(-k)^{-\nu+\frac{1}{2}}}{2\nu}
\frac{h_\ell(k,r)}{w_\ell(kr)}\,,
\label{factorw}
\end{eqnarray}
where $w_\ell$ is the free Jost solution introduced above. With
these definitions,
\begin{equation}
G_\ell(r,r,k) = \frac{h_\ell(k,r)
g_\ell(k,r)}{2\nu g_{\ell}(k,0)}\, .
\label{Green3}
\end{equation}
As we shall show, both $g_{\ell}(k,r)$ and $h_{\ell}(k,r)$ are
well--behaved on the imaginary axis.  Note that the definition of
$h_\ell$ does \emph{not} just remove the free part.\footnote{Removing
the free part would correspond to an {\it ansatz} like
$\phi_\ell\sim w_\ell h_\ell$ as in computations
of functional determinants with Euclidean Green's functions~\cite{Ba93}.}
Instead, it enforces the cancellation of $w_\ell$ in the Green's
function. The factors in eq.~(\ref{factorw}) were chosen to provide
a simple boundary condition on $h_{\ell}(k,r)$ at $r=0$.  After
analytically continuing to $k=it$, $g_{\ell}(it,r)$ obeys
\begin{equation}
g_\ell''(it,r) = 2 t \xi_\ell(t r) g_\ell'(it,r) + \sigma(r) g_\ell(it,r)
\label{ODE1}
\end{equation}
with the boundary conditions
\begin{equation}
\lim_{r\to\infty}g_\ell(it,r) = 1
\qquad {\rm and} \qquad
\lim_{r\to\infty}g_\ell'(it,r) = 0 \,,
\end{equation}
where a prime indicates a derivative with
respect to the radial coordinate $r$.  Using these boundary
conditions, we integrate this differential equation numerically for
$g_{\ell}(it,r)$, starting at $r=\infty$ and proceeding to
$r=0$. Similarly, $h_{\ell}(it,r)$ obeys
\begin{equation}
h_\ell''(it,r) = - 2 t \xi_\ell(tr) h_\ell'(it,r)+
\left[ \sigma(r) - 2 t^2 \left.\frac{d
\xi_\ell(\tau)}{d\tau}\right|_{\tau = t r}\right] h_\ell(it,r) \label{ODE2}
\end{equation}
with the boundary conditions
\begin{equation}
h_\ell(it,0) = 0
\qquad {\rm and} \qquad
h_\ell'(it,0) = 1
\end{equation}
and we integrate numerically from $r=0$ to $r=\infty$. For real $\tau$,
\begin{equation}
\xi_\ell(\tau) \equiv -\frac{d}{d\tau} \ln\left[w_\ell(i\tau)\right]
\label{xi}
\end{equation}
is real with $\lim_{\tau\to\infty}\xi_\ell(\tau)= 1$, so the two
functions $h_\ell(it,r)$ and $g_\ell(it,r)$ are manifestly real. They
are also holomorphic in the upper $k$--plane and, most
importantly, they are bounded according to
\begin{eqnarray}
|g_\ell(k,r)| &\le& \hbox{const}\cr
|h_\ell(k,r)| &\le& \hbox{const} \frac{2\nu r}{1+|k| r}
\label{bounds}
\end{eqnarray}
so that neither $g_{\ell}$ nor $h_\ell$ grows exponentially
during the numerical integration.  Thus the representation of the
partial wave Green's function in terms of $g_{\ell}$ and
$h_{\ell}$ is smooth and numerically tractable on the positive
imaginary axis.  We have avoided oscillating functions or
exponentially growing functions that eventually would cancel against
exponentially decaying functions.

Finally, the computation of the Born series,
eq.~(\ref{state_dens}), is also straightforward in this
formalism.  We expand the solutions to the differential equations
eq.~(\ref{ODE1}) and eq.~(\ref{ODE2}) about the free solutions,
\begin{eqnarray}
g_\ell(it,r)&=&1+g^{(1)}_\ell(it,r)+g^{(2)}_\ell(it,r)+\ldots \cr
h_\ell(it,r)&=&2\nu r I_\nu(tr) K_\nu(tr)+
h^{(1)}_\ell(it,r)+h^{(2)}_\ell(it,r)+\ldots\, ,
\label{expODE}
\end{eqnarray}
where the superscript labels the order of the background potential~${\sigma}$.
The higher order components obey inhomogeneous linear
differential equations with the boundary conditions
\begin{eqnarray}
\lim_{r\to\infty}g^{(j)}_\ell(it,r) =
0\quad&\hbox{and}&\quad\lim_{r\to\infty}
g^{(j)\prime}_\ell(it,r)=0, \cr
h^{(j)}_\ell(it,0) = 0\quad&\hbox{and}&\quad h^{(j)\prime}_\ell(it,0)=0\,.
\label{expbc}
\end{eqnarray}
In these equations ${{\sigma}}$ is the source term for $g^{(1)}$,
${{\sigma}}g^{(1)}$ is the source term for $g^{(2)}$, and so on. We
then obtain the Born series for the local spectral density by substituting
these solutions in the expansion of eq.~(\ref{Green3}) with respect to
the order of the background potential.  Thus we have developed a
computationally robust representation for the Born subtracted local spectral density,
\begin{equation}
\left[ \rho_{\ell}(it,r)\right]_{N} =
\left[ t\frac{h_\ell(it,r) g_\ell(it,r)}
{2\nu g_{\ell}(it,0)}\right]_{N}\, .
\end{equation}

\bigskip
\stepcounter{subsection}
\leftline{\bf 2.3 Feynman Diagram Contribution}
\bigskip

\begin{figure}
\centerline{
  {\epsfig{file=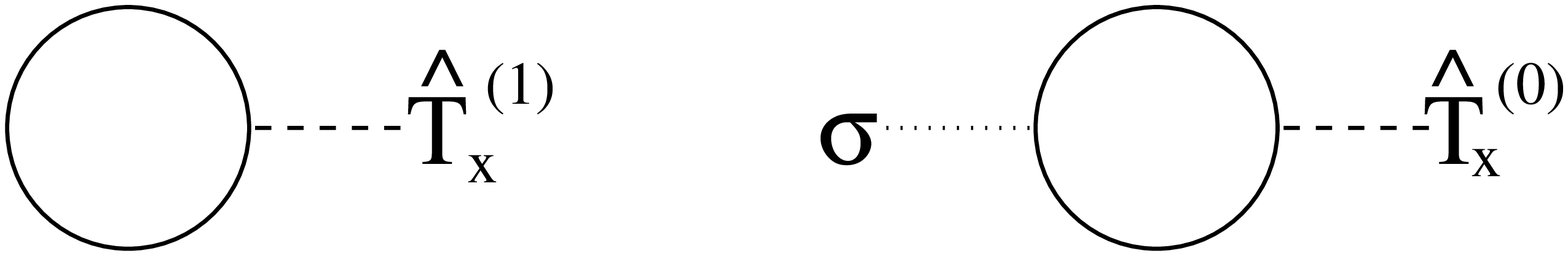, width=10cm}}}
\caption{\small First order Feynman diagrams contributing
to the energy density $\langle \hat{T}_{00}(x)\rangle$.  The solid
circle represents the $\phi$ loop and the operator insertions from
eq.~(\ref{defoperatorT}) are denoted by $\hat{T}_x^{(0,1)}$. In the
right panel, $\sigma$ is the local insertion of the background field
from expanding the propagator, {\it cf.}~eq.~(\ref{twopoint1}).}
\label{diagrams}
\end{figure}

In this subsection we develop methods to compute the contribution
from the Feynman diagrams and counterterms,
$\sum\limits_{i=0}^{N} \epsilon_{\rm FD}^{(i)}(r) + \epsilon_{\rm CT}(r)$,
in eq.~(\ref{fundamental}).  To one--loop order, the Feynman diagrams
of interest are generated by the expansion of
\begin{equation}
\langle0|\hat{T}_{00}(x)|0\rangle \sim\frac{i}{2} {\rm Tr}\,
\left[\hat{T}_x \left(-\partial^2-m^2 -\sigma\right)^{-1}\right]
\label{t00matrix}
\end{equation}
to order $N$ in the background $\sigma$.  Here $\hat T_{x}$ is the
coordinate space operator corresponding to the insertion of the energy
density defined by eq.~(\ref{H_dense}) at the spacetime point $x$, and
the trace includes space--time integration.  Since we are doing
ordinary perturbation theory, we take the matrix elements with respect
to the trivial vacuum, which is annihilated by the plane wave
annihilation operators.  The energy density operator has pieces of
order $\sigma^0$ and $\sigma^1$,
\begin{equation}
\hat{T}_{00}=\frac{1}{2}\left[\dot{\phi}^2+
\left(\vec{\nabla} \phi\right)^2+m^2\phi^2+\sigma\phi^2\right]
=\frac{1}{2}\int d^dy \phi(y)
\left(\hat{T}^{(0)}_x+\hat{T}^{(1)}_x\right) \phi(y)
\end{equation}
where $d=n+1$ is the number of space--time dimensions.  We have
separated the different orders in the external background field,
\begin{eqnarray}
\hat{T}^{(0)}_x &=& \left[\ldert_t \delta(x-\hat{y}) \rdert_t + \lder
\delta(x-\hat{y}) \rder + m^2\delta(x-\hat{y})\right] \cr
\hat{T}^{(1)}_x &=& \sigma(x)\delta(x-\hat{y})\, .
\label{defoperatorT}
\end{eqnarray}
The $\delta$-functions are understood as $d$--dimensional.  The
arrows denote the direction in which the derivatives act when
substituting this representation into the functional trace in
eq.~(\ref{t00matrix}) and $\hat{y}$ is the coordinate space position
operator, with $\hat{y}|\xi\rangle=\xi|\xi\rangle$.  However, it is more
convenient to perform the computation in momentum space. The
relevant matrix elements are
\begin{eqnarray}
\langle k^\prime | \hat{T}^{(0)}_x | k \rangle
&=&e^{i(k^\prime-k)x}
\left[k^{0\prime} k^{0} +\vec{k}^\prime\cdot\vec{k} +m ^2\right] \cr
\langle k^\prime | \hat{T}^{(1)}_x | k \rangle
&=&\sigma(x)e^{i(k^\prime-k)x}\, .
\label{t00operator}
\end{eqnarray}
Here we will explicitly consider the contributions to $ \langle
\hat{T}_{00}\rangle$ that are linear in $\sigma$ because these are
the only contributions required in the examples studied in
Sections~3 and~4. The first piece comes directly from
$\hat{T}_x^{(1)}$ as shown in the left panel of Fig.~\ref{diagrams},
\begin{equation}
\frac{i}{2}{\rm Tr}\left(\frac{1}{-\partial^2-m^2} \hat{T}^{(1)}_x \right)=
\frac{i}{2}\sigma(x)\int \frac{d^dk}{(2\pi)^d}\frac{1}{k^2-m^2}\, .
\label{direct}
\end{equation}
This local contribution is ultraviolet divergent for $d\ge 2$.  It is
canceled identically by the counterterm in the no--tadpole
renormalization scheme, since the right--hand side of
eq.~(\ref{direct}) equals minus the counterterm Lagrangian evaluated
for the background field $\sigma$. For further discussion of this
renormalization condition and its consequences, see Ref.~\cite{method1}.
The second contribution at order $\sigma$ originates from
$\hat{T}^{(0)}_x$ and the first--order expansion of the propagator as
shown in Fig.~\ref{diagrams},
\begin{equation}
\frac{i}{2}{\rm Tr}\left(\frac{1}{-\partial^2-m^2}
\hat{T}^{(0)}_x \frac{1}{-\partial^2-m^2}\sigma\right)\,.
\label{twopoint1}
\end{equation}
For this diagram we find
\begin{eqnarray}
&&\frac{i}{2}\int \frac{d^dk}{(2\pi)^d} \int
\frac{d^dk^\prime}{(2\pi)^d} \frac{1}{k^2-m^2}
e^{i(k^\prime-k)x} \left(\tilde k\cdot k^\prime +m^2\right)
\frac{1}{k^{\prime2}-m^2}\tilde\sigma(k-k^\prime) \cr
&& = \frac{i}{2}\int \frac{d^dq}{(2\pi)^d}
\tilde\sigma(q) e^{-iqx} \int_0^1 d\zeta \int
\frac{d^dl}{(2\pi)^d} \frac{\tilde{l}\cdot
l+m^2-\zeta(1-\zeta)\tilde{q}\cdot q}
{\left[l^2-m^2+\zeta(1-\zeta)q^2\right]^2}\, ,
\label{twopoint2}
\end{eqnarray}
where $\tilde\sigma(q)=2\pi\delta(q^{0})\tilde\sigma(\vec q)$ is the Fourier
transform of the (time independent) background field.  We have defined
$\tilde{p} = (p^{0},-\vec{p})$ for the Lorentz-vector $p=(p^{0},\vec{p})$.
The leading ultraviolet divergences cancel in eq.~(\ref{twopoint2})
between the temporal and spatial components.  We thus obtain the
expression for the Feynman diagram and counterterm contribution
through first order in $\sigma$,
\begin{eqnarray}
\epsilon_{\rm FD}^{(1)}(r) + \epsilon_{\rm CT}(r)&=&
\frac{2\pi^{\frac{n}{2}}r^{n-1}}{\Gamma(\frac{n}{2})}
\frac{i}{2}{\rm Tr}\left(\frac{1}{-\partial^2-m^2}
\hat{T}^{(0)}_x
\frac{1}{-\partial^2-m^2}\sigma\right) \label{final}  \cr
&=& \frac{2\pi^{\frac{n}{2}} r^{n-1}}{\Gamma(\frac{n}{2})}
\frac{\Gamma\left(2-\frac{d}{2}\right)}{(4\pi)^{d/2}} \int
\frac{d^nq}{(2\pi)^n} \tilde\sigma(\vec{q})
e^{i\vec{q}\cdot\vec{x}}\int_0^1 d\zeta
\frac{\zeta(1-\zeta){\vec q\,}^2 }
{\left[m^2+\zeta(1-\zeta){\vec q\,}^2\right]^{2-d/2}}\, .
\end{eqnarray}

We note that this piece does not contribute to the total energy
because it vanishes when integrated over space.  The extension to
higher order Feynman diagrams in $\sigma$ is straightforward.  For
example, the contribution ${\cal O}(\sigma^2)$ is obtained from the two terms
\begin{equation}
\frac{i}{2}{\rm Tr}\,\left(\hat{T}_x^{(1)}\frac{1}{\partial^2+m^2}
\sigma\frac{1}{\partial^2+m^2}\right)
-\frac{i}{4}{\rm Tr}\,\left(\hat{T}_x^{(0)}\frac{1}{\partial^2+m^2}
\sigma\frac{1}{\partial^2+m^2}\sigma\frac{1}{\partial^2+m^2}\right)\,.
\end{equation}
We use the matrix elements, eq.~(\ref{t00operator}), to write this in
terms of ordinary loop integrals
\begin{eqnarray}
&&\frac{i}{2}\sigma(x)
\int \frac{d^dq}{(2\pi)^d}\tilde\sigma(q)e^{iqx}
\int_0^1 d\zeta \frac{d^dl}{(2\pi)^d}
\frac{1}{\left[l^2-m^2+\zeta(1-\zeta)q^2\right]^2}
\cr &&
+\frac{i}{2}\int \frac{d^dq}{(2\pi)^d}\int \frac{d^dp}{(2\pi)^d}
\tilde\sigma(q-p)\tilde\sigma(p)e^{-iqx}
\cr && \times \int_0^1d\zeta\int_0^{1-\zeta}d\eta
\int\frac{d^dl}{(2\pi)^d} \frac{\tilde{l}\cdot l
+\left[\zeta\tilde{q}+\eta\tilde{p}\right]\cdot
\left[(1-\zeta)q-\eta p\right]}
{\left[l^2-m^2+\zeta(1-\zeta)q^2+\eta(1-\eta)p^2\right]^3}\,.
\end{eqnarray}
As a consequence of eq.~(\ref{t00operator}), the term involving
$\hat{T}_x^{(1)}$ will always carry an explicit factor of $\sigma(x)$.

\bigskip

\stepcounter{subsection}
\leftline{\bf 2.4 Total Energy}
\bigskip

\noindent
We can obtain the total energy simply by integrating the energy density in
eq.~(\ref{fundamental}),
\begin{equation}
E[\sigma] = \int_0^\infty \epsilon(r) dr \,.
\label{denstot}
\end{equation}
Since both the $t$ integral and the sum over channels in 
eq.~(\ref{fundamental}) are absolutely convergent, we can
interchange the order of integration and obtain
\begin{equation}
E[\sigma] = - \sum\limits_\ell N_\ell \int_m^\infty \frac{dt}{\pi}\,
\sqrt{t^2-m^2}\int_0^\infty dr \,\left[\rho_\ell(it,r)\right]_N + 
\sum_{i=1}^N E_{\rm FD}^{(i)} + E_{\rm CT} \,,
\label{totimag}
\end{equation}
where the total derivative term has integrated to zero.  As explained
in the previous subsection, we have to subtract a sufficient number of
Born approximations to the local spectral density $\rho_\ell(it,r)$ to render the
$t$ integral convergent.  These subtractions are then added
back in as the contribution to the total energy from the Feynman diagrams
\begin{equation}
\sum_{i=1}^N E_{\rm FD}^{(i)} = \sum_{i=1}^N \int_0^\infty dr\,
\epsilon_{\rm FD}^{(i)}(r)\,,
\label{ballack}
\end{equation} 
which we then combine with the contribution from the counterterms
\begin{equation}
E_{\rm CT} = \int_0^\infty dr\,\epsilon_{\rm CT}(r)
\end{equation}
to obtain a finite result.  In practice, these terms can be more
efficiently computed directly from the perturbation series of the
total energy.  For the first term in eq.~(\ref{totimag}), we employ
the relation
\begin{equation}
2 \int_0^\infty dr\,\left[\rho_\ell(k,r)\right]_0 = 
\frac{2 k}{i} \int_0^\infty dr\,\left[G_\ell(r,r,k) - G^{(0)}_\ell(r,r,k)
\right] = i \frac{d}{dk} \ln F_\ell(k)
\label{app1}
\end{equation}
which we prove in Appendix A for all $k$ in the upper half--plane,
$\mathsf{Im}\,k > 0$. Using eq.~(\ref{app1}) on the imaginary axis
$k=it$, we find in particular
\begin{equation}
2 \int_0^\infty dr\, \left[\rho(it,r)\right]_0 = 
\frac{d}{dt} \ln F_\ell(it) = \frac{d}{dt} \ln \,\lim_{r\to 0}
\frac{f_\ell(it,r)}{w_\ell(itr)} = \frac{d}{dt} \ln g_\ell(it,0)\,.
\label{app2}
\end{equation}
Finally, we introduce the real function $\beta_\ell(t,r) = \ln g_\ell(it,r)$ 
and write
\begin{eqnarray}
E[\sigma]&=&-\sum_\ell N_\ell \int_m^\infty \frac{dt}{2\pi}
\sqrt{t^2-m^2} \frac{d}{dt}
\left[ \beta_\ell(t,0) \right]_{N}
+ \sum_{i=1}^{N} E_{\rm FD}^{(i)} + E_{\rm CT} \cr
&=&\sum_\ell N_\ell \int_m^\infty \frac{dt}{2\pi}
\frac{t}{\sqrt{t^2-m^2}}
\left[ \beta_\ell(t,0) \right]_{N} + \sum_{i=1}^{N} E_{\rm FD}^{(i)}
+ E_{\rm CT}\,,
\label{evac3}
\end{eqnarray}
where $\beta_\ell$ is determined by the differential equation
\begin{equation}
-\beta_\ell^{\prime\prime}(t,r)-\left[\beta^\prime_\ell(t,r)\right]^2
+2t\xi_\ell(tr)\beta^\prime_\ell(k,r)+\sigma(r) = 0
\label{betadeq}
\end{equation}
with the boundary conditions 
$\lim_{r\to\infty}\beta(t,r)=\lim_{r\to\infty}\beta^\prime(t,r)=0$.
To compute $\left[ \beta_\ell(t,0) \right]_{N}$, we must subtract the first
$N$ Born terms, which we compute by iterating the differential
equation (\ref{betadeq}) according to the expansion of $\beta_\ell(t,r)$
in powers of the interaction~$\sigma$. 

The formulation (\ref{evac3}) is well suited for numerical evaluation
and we will use it exclusively to compute the total energy.
To make contact with previous work \cite{method1}, however, it is
instructive to return to the real axis.  We rotate the contour in
eq.~(\ref{evac3}) to obtain an integral over the whole real axis,
picking up the discrete contributions from the bound state poles.
When $k$ approaches the real axis from above, we have
\begin{equation}
i \frac{d}{dk } \ln F_\ell(k) = 
i\frac{d}{dk}\ln |F_\ell(k)| + \frac{d\delta_\ell(k)}{dk}
\label{beckenbauer}
\end{equation}
where $\delta_\ell(k)$ is the phase shift.  The first term on the
right--hand side is odd in $k$ and does not contribute to the
integral.  Thus we have
\begin{equation}
E[\sigma] = \sum\limits_\ell N_\ell\left[
\int_0^\infty \frac{dk}{2\pi}\sqrt{k^2 + m^2} \frac{d}{dk}
[\delta_\ell(k)]_N + \frac{1}{2}\sum_j \omega_j\right]
+ \sum_{i=1}^N E_{\rm FD}^{(i)} +  E_{\rm CT}
\label{previousdelta}
\end{equation}
which was used in \cite{method1}.  To understand the physical
motivation for eq.~(\ref{previousdelta}), we introduce the difference
between the free and interacting density of states,
\begin{equation}
\frac{2k}{\pi} \mathsf{Im} \int_0^\infty \left(
G_\ell(r,r,k+i\epsilon)-G_\ell^{(0)}(r,r,k+i\epsilon)\right) dr =
\rho_\ell(k) - \rho_\ell^{(0)}(k) \,,
\label{dos}
\end{equation}
which from eq.~(\ref{app1}) and eq.~(\ref{beckenbauer}) is related to the
phase shift by
\begin{equation}
\rho_\ell(k) - \rho_\ell^{(0)}(k) = \frac{1}{\pi}\frac{d\delta_\ell}{dk} \,.
\label{realaxis}
\end{equation}
Thus we can write the energy as 
\begin{equation}
E[\sigma] = \sum\limits_\ell N_\ell\left[
\int_0^\infty \frac{dk}{2\pi}\sqrt{k^2 + m^2} [\rho_\ell(k)]_N +
\frac{1}{2}\sum_j \omega_j
\right] + \sum_{i=1}^N E_{\rm FD}^{(i)} + 
E_{\rm CT} \, ,
\label{previous}
\end{equation}
and see that it is simply the sum of the zero--point energies
$\fract{1}{2} \omega$, including both discrete contributions from the
bound states and an integral over the continuum, weighted by the
density of states.

We can also see this relation directly by integrating
eq.~(\ref{ebare}) over $r$, with $N$ Born terms subtracted and added
back in as diagrams, 
\begin{equation}
E[\sigma] = \sum_\ell N_\ell\left[
\int_{0+i\epsilon}^{\infty+i\epsilon} \frac{dk}{\pi}\sqrt{k^2 + m^2}
\int_0^\infty dr \left[\psi_\ell^*(k,r)\psi_\ell(k,r)\right]_N +
\sum_j\frac{\omega_j}{2}\right]
 + \sum_{i=1}^N E_{\rm FD}^{(i)} + E_{\rm CT} \,,
\label{intdensity}
\end{equation}
where the total derivative term has integrated to zero and the bound
state wavefunctions have integrated to unity.  We can then use
eq.~(\ref{normtophase}) to see that eq.~(\ref{intdensity}) is
equivalent to eq.~(\ref{previous}).  While this relation is useful for
building an intuitive understanding of the expressions for the total
energy and the energy density, eq.~(\ref{intdensity}) is of no
practical use in numerical calculations; it only converges
because eq.~(\ref{dos}) has introduced the standard $i\epsilon$
prescription to control the oscillations of the wavefunctions at
spatial infinity.  Any direct numerical calculation using
eq.~(\ref{intdensity}) will be hopelessly obscured by these oscillations.

\bigskip
\stepcounter{section}
\leftline{\large\bf 3. Examples in One Space Dimension}
\bigskip

To illustrate our method, we apply it first
to simple Casimir problems in one space dimension.  There
are only two channels (symmetric and antisymmetric) in this case.  The
boundary condition in the symmetric channel,
$\frac{d\phi}{dx}|_{x=0}=0$, requires slight modifications of the
formalism developed so far.  We study two examples:  a background
field consisting of $\delta$-functions, which allows us to make
contact with treatments already in the literature
\cite{MT,MiltonReview}, and a Gau{\ss}ian background, to demonstrate
the efficient use of our methods for numerical computation.

\bigskip

\stepcounter{subsection}
\leftline{\bf 3.1 Green's Function}
\bigskip

In one spatial dimension, the S--matrix for a symmetric potential is
diagonalized in a basis of symmetric and antisymmetric
wavefunctions. In the antisymmetric channel, we can proceed with the
general techniques described in Section 2.  From eqs.~(\ref{ODE1})
and~(\ref{ODE2}) we compute $g(it,x)$ and $h_{-}(it,x)$ with $\xi=1$.
Note that $g(it,x)$ is the same in both channels because the channels
only differ by the boundary conditions at the origin, which do
not affect $g$.  At equal spatial arguments we
have 
\begin{equation}
G_-(x,x,it)=\frac{h_-(it,x)g(it,x)}{g(it,0)}\,.
\label{G1das}
\end{equation}
However, the boundary conditions in the symmetric channel,
\begin{equation}
\phi_+(0)=1\qquad {\rm and} \qquad
\phi_{+}^\prime(0)=0 \,,
\label{bcsym}
\end{equation}
are not in the class defined by eq.~(\ref{bcregular}).
Constructing the Jost solution
$f(k,x)$ with the boundary condition $\lim_{x\to\infty}f(k,x)
e^{-ikx}=1$ and working out the Wronskian for $\phi_+$ and $f$
yields the Green's function in the symmetric channel
\begin{equation}
G_+(x,y,k)=-\frac{\phi_+(k,x_<)f(k,x_>)}{f^\prime(k,0)}\,,
\label{Grennsym}
\end{equation}
where $f^\prime(k,0) = \partial_x f(k,x) |_{x=0}$.
By symmetry we have $G_+(-x,-y,k)=G_+(x,y,k)$.  We factor out the
asymptotic behavior,
\begin{equation}
\phi_+(k,x)=e^{-ikx}h_+(k,x)
\label{paraphiP}
\end{equation}
and obtain
\begin{equation}
-h_+^{\prime\prime}(it,x)-2th_+^{\prime}(it,x)+\sigma(x)h_+(it,x)=0\, .
\label{deqhP}
\end{equation}
which is the same differential equation obeyed by
$h_{-}(it,x)$. The boundary conditions in eq.~(\ref{bcsym})
then become
\begin{equation}
h_+(it,0)=1 \qquad {\rm and} \qquad h_+^{\prime}(it,0)=-t\, .
\label{bcsym1}
\end{equation}
With this solution, we can now construct the Green's function in
the symmetric channel.  To compute densities it again suffices to
consider $x=y$, where we have
\begin{equation}
G_+(x,x,it)=\frac{g(it,x)h_+(it,x)}{tg(it,0)-g^\prime(it,0)}
\label{Grennsym1}
\end{equation}
and the complete Green's function is the sum $G=G_{+}+G_{-}$, giving
\begin{equation}
\rho(it,x)= t \frac{g(it,x)h_-(it,x)}{g(it,0)} +
\frac{g(it,x)h_+(it,x)}{g(it,0)-g^\prime(it,0)/t}\, .
\label{dens1d}
\end{equation}
The Born series is then obtained by iteration
of the differential equations (\ref{ODE1}), (\ref{ODE2}), and (\ref{deqhP}).

\bigskip
\stepcounter{subsection}
\leftline{\bf 3.2 Total Energy}
\bigskip

The contribution of the symmetric channel to the total energy
cannot be cast in the general form of eq.~(\ref{evac3})
because, as for the Green's function, the boundary condition for
the corresponding wavefunction is not of the type of
eq.~(\ref{bcregular}).  Since the derivative of ${\phi_{+}}$ vanishes
at $x=0$, the Jost function becomes
\begin{equation}
F_+(k)=\lim_{x\to0}\frac{f^\prime(k,x)}{ik}
=g(k,0)+\frac{g^\prime(k,0)}{ik}
\label{jostsym}
\end{equation}
which is normalized so that $F_+(k)=1$ when $\sigma\equiv 0$.
Otherwise, the same analyticity arguments apply as in
channels whose wavefunctions vanish at $r=0$.  In particular, the $k$
integration contour can be closed around the discontinuity along
the positive imaginary axis. We obtain
\begin{equation}
E[\sigma]=\int_m^\infty\frac{dt}{2\pi}\frac{t}{\sqrt{t^2-m^2}}
\left[\ln g(it,0)
+\ln\left(g(it,0)-\frac{1}{t}g^\prime(it,0) \right)\right]_{1}
+ E_{\rm FD}^{(1)} + E_{\rm CT}\, ,
\label{evac1d}
\end{equation}
where as before the subscript indicates that the first
Born term has been removed from the expression in brackets.
One subtraction suffices in one dimension.  By our no--tadpole
renormalization condition, the counterterm cancels the tadpole diagram
identically, so that $E_{\rm FD}^{(1)} + E_{\rm CT}\equiv0$.

\bigskip
\stepcounter{subsection}
\leftline{\bf 3.3 Delta--Function Background}
\bigskip

We first consider the case of two $\delta$-functions separated by a
distance $2a$,
\begin{equation}
\sigma_\delta(x)=\lambda\left[\delta(x+a)+\delta(x-a)\right]\, .
\label{delpot}
\end{equation}
At each $\delta$-function, the derivative of the field
$\phi$ jumps by $\lambda\phi$.  For modes with
$k\ll\lambda$, the wavefunction $\psi(k,x)$ vanishes
at $\pm a$, so that the limit $\lambda\to\infty$ gives a Dirichlet
boundary condition at $\pm a$. Since the counterterms are local
functionals of the background field, they are identically zero in
regions where the background fields vanish.  Hence in a renormalizable
theory, local observables such as the energy density cannot have
ultraviolet divergences in these regions, as we will see explicitly below. 

By integration of the respective differential equations we obtain
\begin{eqnarray}
g(it,x) &=& \left\{ \begin{array}{l@{\quad\quad}l}
1 & |x|>a \\ 
\displaystyle 1+\frac{\lambda}{2t}\left[1-e^{2t(x-a)}\right] & 0\le |x|<a
\end{array}\right.
\label{2delg}\nonumber \\[1mm]
h_+(it,x)&=& \frac{1}{2}\left\{\begin{array}{l@{\quad\quad}l}
\displaystyle 1+e^{-2tx}+\frac{\lambda}{2t} \left[1+e^{-2ta}-e^{-2tx}
-e^{-2t(x-a)}\right] & |x|>a \\
\displaystyle 1+e^{-2tx} & 0\le |x|<a \end{array}\right.
\label{2delh}
\\[1mm]
h_-(it,x)&=& \frac{1}{2t}\left\{\begin{array}{l@{\quad\quad}l}
\displaystyle 1-e^{-2tx}+\frac{\lambda}{2t}\left[1-e^{-2ta}+e^{-2tx}
-e^{-2t(x-a)}\right] & |x|>a \\
\displaystyle 1-e^{-2tx} & 0\le |x|<a \end{array}\right.
\nonumber
\end{eqnarray}
It is clear that $g$, $h_{+}$, and $h_{-}$ are all
well--behaved for $t>0$.  We substitute $g(it,0)$ and its
derivative into eq.~(\ref{evac1d}) and find the total energy
\begin{equation}
E_2(a,\lambda)=\int_m^\infty\frac{dt}{2\pi}\frac{1}{\sqrt{t^2-m^2}}
\left\{t \ln\left[1+\frac{\lambda}{t}
+\frac{\lambda^2}{4t^2}\left(1-e^{-4ta}\right)\right]
-\lambda\right\}\,.
\label{evac2del}
\end{equation}
The last term in curly brackets is the $N=1$ Born subtraction. 
Carrying out a similar calculation for a single
$\delta$-function, we obtain
\begin{eqnarray}
E_1(\lambda)&=&\int_m^\infty\frac{dt}{2\pi} \frac{t \ln
\left[1+\frac{\lambda}{2t}\right]-\frac{\lambda}{2}}{\sqrt{t^2-m^2}} 
\\[1mm]
&=&
\frac{\lambda-m\pi}{4\pi}+
\frac{m}{2\pi} \left\{\begin{array}{l@{\quad\quad}l}
\sqrt{1-\frac{\lambda^2}{4m^2}}
\arccos \frac{\lambda}{2m} & \lambda\le 2m \\[2mm]
\sqrt{\frac{\lambda^2}{4m^2}-1} \ln
\left(\frac{\lambda}{2m}-\sqrt{\frac{\lambda^2}{4m^2}-1}\right)
& \lambda>2m\,.\end{array}\right.
\label{evac1del}
\end{eqnarray}
From eq.~(\ref{evac1del}) we can verify that $E_{1}$ and
$E_{2}$ obey the consistency conditions
$\lim\limits_{a\to\infty}E_{2}(a,\lambda) = 2 E_{1}(\lambda)$ and 
$\lim\limits_{a\to 0} E_{2}(a,\lambda)= E_{1}(2\lambda)$.

We observe that in the limit of large $\lambda$ the total
energy associated with a single $\delta$-function approaches
minus infinity as $-\lambda \ln\lambda$, which cannot be
canceled by any available counterterm.  Thus in this limit
the energy is infinite.  \emph{In this sense the Dirichlet--Casimir
problem is not well defined in the context of renormalizable
quantum field theory.} However, we can compute the force between
two $\delta$-functions for finite $\lambda$ and subsequently
take the $\lambda\to\infty$ limit,
\begin{equation}
K(a)=-\lim_{\lambda\to\infty}
\frac{\partial E_2(a,\lambda)}{\partial (2a)}
=-\int_m^\infty \frac{dt}{\pi}
\frac{t^2}{\sqrt{t^2-m^2} \left(e^{4ta}-1\right)}\, .
\label{casforce}
\end{equation}
which is equal to the result found using boundary conditions
\cite{MT}.  The massless limit does not exist for the vacuum
polarization energy in eq.~(\ref{evac1del}), even for finite
$\lambda$. This is to be expected since massless scalar
field theories are infrared divergent in $1+1$ dimensions. However, it
exists for the force, yielding the well--known result
\begin{equation}
\lim_{m\to0}K(a)=-\frac{\pi}{96a^2}\, .
\label{casforce1}
\end{equation}

Next we turn to the energy density. In this Section, we consider the
energy density $\epsilon(x)  = \langle T_{00} \rangle$ without
factoring out the surface area factor as we did in the previous
Section (which would just give a factor of $2$ here).  The resulting
energy density $\epsilon(x)$ is normalized such
that $\int_{-\infty}^\infty \epsilon(x) dx$ gives the total energy.
It is convenient to decompose the energy density into
$\bar\epsilon(x)$, the integral over the subtracted local spectral density,
and $\epsilon_{\rm FD}^{(1)}(x) + \epsilon_{\rm CT}(x)$, the
contribution of the Feynman diagram plus counterterms.

First we study $\bar\epsilon(x)$.  From the radial functions in
eq.~(\ref{2delg}), we find the local spectral density as a function of $x$
and $k=it$,
\begin{eqnarray}
\left[\rho(it,x)\right]_{0}=\frac{\lambda}
{\lambda^2-(2t+\lambda)^2e^{4ta}} \cases{ \displaystyle
\left[2t-\lambda+(2t+\lambda) e^{4ta}\right] 
e^{2t(a-|x|)}& $|x|>a$\cr &\cr \displaystyle 2\left[(2t+\lambda)
e^{2ta} \cosh(2tx) - \lambda\right] & $0\le |x|<a$\,,}
\label{dens0}
\end{eqnarray}
where only the free local spectral density has been subtracted.  For
$|x|\ne a$ the local spectral density decays exponentially as $t$
increases, so it yields a finite energy density without
any subtractions, as we expected since the counterterm
vanishes away from $x=\pm a$.

To prepare for more general problems involving smooth potentials,
where the energy will diverge at all $x$, we carry out the
full Born subtraction procedure even though in this case the Born
subtraction and the diagram contribution are both finite and equal for all
$|x|\neq a$, and so their contributions cancel.  We will thus obtain the
energy density everywhere  including the points $x=\pm a$.
The first Born approximation is given by
\begin{eqnarray}
{\rho}^{(1)}(it,x)=-\frac{\lambda}{2t}
\cases{
\displaystyle
e^{-2t(|x|-a)}+e^{-2t(|x|+a)},& $|x|>a$\cr
&\cr
\displaystyle
e^{2t(|x|-a)}+e^{-2t(|x|+a)},& $0\le |x|<a$}
\label{dens0o1}
\end{eqnarray}
and we obtain
\begin{equation}
\bar\epsilon(x,a,\lambda)=
\int_m^\infty \frac{dt}{\pi} \frac{\eta(it,x)}{\sqrt{t^2-m^2}}
\label{ebs}
\end{equation}
where
\begin{eqnarray}
\eta(it,x)&=&\frac{\lambda^2}{2}
\frac{2t-\lambda+(2t+\lambda)e^{4ta}}
{\lambda^2-(2t+\lambda)^2e^{4ta}} \delta(|x|-a) \cr
&+&\frac{m^2\lambda^2/t}{(2t+\lambda)^2e^{4ta}-\lambda^2}
\cases{\fract{1}{2}\left[6t-\lambda e^{-4ta}+
(2t+\lambda)e^{4ta}\right]e^{2t(a-|x|)}& $|x|>a$\cr &\cr
\frac{2t}{m^2}(m^2-t^2)
+\left[(2t+\lambda)e^{2ta}-\lambda e^{-2ta}\right]
\cosh(2tx)  & $0\le |x|<a$\,.}
\label{rhobs}
\end{eqnarray}

For the Feynman diagram and the counterterm, we use
eq.~(\ref{final}) and split off a $\delta$-function contribution,
\begin{eqnarray}
\epsilon_{\rm FD}^{(1)}(x,a,\lambda)+\epsilon_{\rm
CT}(x,a,\lambda)&=&\frac{\lambda}{2\pi}\delta(|x|-a) \cr
&-&\frac{\lambda}{\pi^2}\int_0^\infty dq \cos(qa) \cos(qx)
\int_0^1 d\zeta \frac{m^2}{m^2+\zeta(1-\zeta)q^2} \,.
\end{eqnarray}
We recast the second term in a form which makes its identification
with the first Born contribution, eq.~(\ref{dens0o1}), manifest,
\begin{eqnarray}
\epsilon_{\rm FD}^{(1)}(x,a,\lambda)&+&\epsilon_{\rm
CT}(x,a,\lambda)=\frac{\lambda}{2\pi}\delta(|x|-a)
-\frac{4m^2\lambda}{\pi^2} \int_0^\infty dq
\frac{\cos(qa) \cos(qx) } {q\sqrt{q^2+4m^2}}
\arsinh \frac{q}{2m} \cr
&=&\frac{\lambda}{2\pi}\delta(|x|-a)
-\frac{m^2\lambda}{2\pi}\int_m^\infty
\frac{dt}{t\sqrt{t^2-m^2}} \left[e^{-2|x-a|t}+
e^{-2|x+a|t}\right] \,,
\label{efd}
\end{eqnarray}
which confirms that our Feynman diagram representation of
the $\cal{O}(\lambda)$ contribution to \linebreak 
$\epsilon(x,a,\lambda)$ agrees
with the $k$ integral representation for values of $x$ where the
counterterms vanish.

\begin{figure}
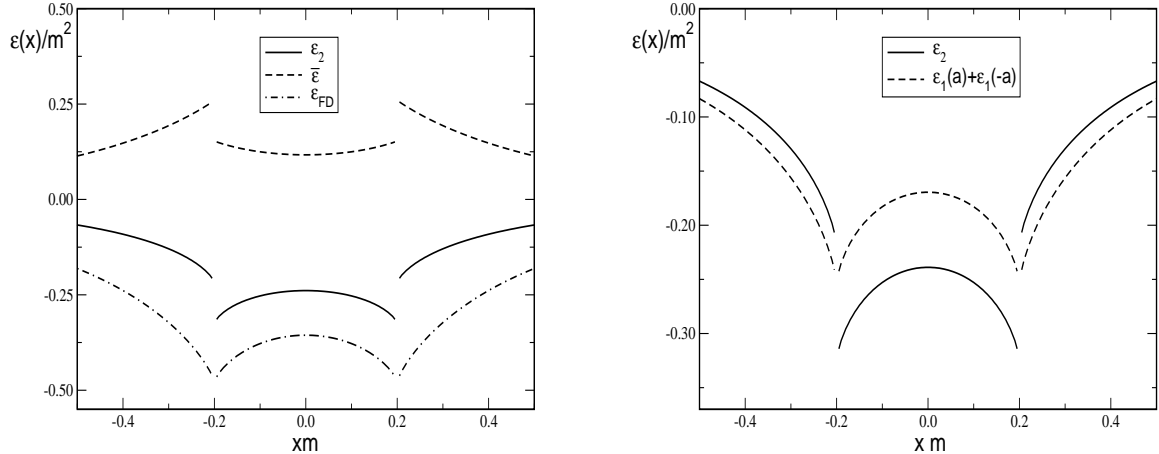

\centerline{ \epsfig{file=dens1.eps,height=6.0cm,width=7cm}
\hspace{1cm} \epsfig{file=dens2.eps,height=6.0cm,width=7cm} }
\caption{\small Left panel: The total energy density
$\epsilon_2(x,a,\lambda)$ and the contributions
$\bar{\epsilon}(x,a,\lambda)$ and $\epsilon_{\rm FD}^{(1)}(x,a,\lambda)$,
defined in eqs.~(\protect\ref{ebs}) and~(\protect\ref{efd}).  Note
that the displayed energy densities are supplemented by
$\delta$-function contributions located at  $x=a$ and $x=-a$.
Right panel:  The total density $\epsilon_{2}(x,a,\lambda)$ compared
to the sum $\epsilon_{1}(x-a,\lambda) + \epsilon_{1}(x+a,\lambda)$ of
the energy densities of two isolated $\delta$-functions.
The parameters are $\lambda=3m$ and $a=0.2/m$.}
\label{fig_ed1}
\end{figure}
In Fig.~\ref{fig_ed1} we display the energy
density for two delta functions separated by $2a$,
$\epsilon_{2}(x,a,\lambda)$ as functions of $x$.  Note that there are
$\delta$-function contributions to $\epsilon_{2}(x,a,\lambda)$ at
$x=\pm a$ not drawn in the figure.  For reference we also
show the sum of the energy densities for delta functions at $x=a$ and
$x=-a$ separately, which we denote $\epsilon_{1}(x\mp a, \lambda)$.
The difference between $\epsilon_{2}(x,a,\lambda)$ and the sum
$\epsilon_{1}(x-a,\lambda)+\epsilon_{1}(x+a,\lambda)$ will result in a
force between the two points.

As $\lambda\to\infty$, we expect to obtain the same energy density as
if we had imposed the boundary conditions $\phi(-a)=\phi(a)=0$ from
the outset. No subtraction is necessary away from $x=\pm a$, so we can
study the energy density in this region directly from
$\left[\rho\right]_{0}$, given in eq.~(\ref{dens0}),
\begin{equation}
\lim_{\lambda\to\infty}\epsilon_2(x)=-\frac{1}{\pi}
\int_m^\infty \frac{dt}{\sqrt{t^2-m^2}}
\frac{t^2-m^2+m^2e^{2ta}\cosh(2tx)}
{e^{4at}-1} \quad \hbox{for}\quad |x|<a \,.
\label{epsinfty}
\end{equation}
To compare with Ref.~\cite{MT}, we separate the
$x$--dependent part and transform the $t$ integral into a
sum over the poles of the function $e^{2ta}/(e^{4at}-1)=1/(2 \sinh
2ta)$ at $t=in\pi/2a$,
\begin{equation}
\lim_{\lambda\to\infty}\epsilon_2(x)=-\frac{1}{\pi} \int_m^\infty
\frac{dt}{\sqrt{t^2-m^2}} \frac{t^2-m^2}{
e^{4at}-1}-\frac{m}{8a} -\frac{m^2}{4a}\sum_{n=1}
\frac{\cos\left[\frac{n\pi}{a}(|x|-a)\right]} {\sqrt{\left(\frac{n
\pi}{2a}\right)^2+m^2}} \quad \hbox{for}\quad |x|<a\,.
\label{epsinftyMT}
\end{equation}
This expression agrees with the Dirichlet result given in Ref.~\cite{MT}.

\begin{figure}
\centerline{
\epsfig{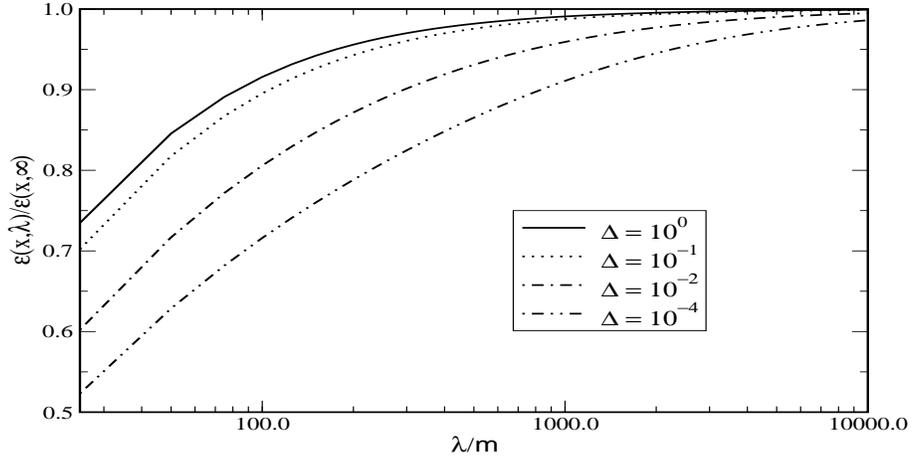}}
\caption{\small Ratio of the energy density for two $\delta$-functions with
finite $\lambda$ to the energy density for $\lambda\to\infty$
with $a=0.2/m$, plotted as a function of $\lambda$ for several values
of $\Delta = (a-x)/a$.}
\label{fig_ed3}
\end{figure}

This simple example allows us to study the nature of the divergence in
the Dirichlet Casimir energy in the limit $\lambda\to\infty$.  For any
fixed $x$ away from the $\delta$-functions, the energy density reaches
a finite limit as $\lambda\to\infty$, but the limit is approached
nonuniformly.  This can be seen clearly in Fig.~\ref{fig_ed3}, where
we display the energy density for large couplings and compare it to
the limiting function, eq.~(\ref{epsinfty}). The closer $x$ approaches
to $a$, the slower the energy density converges to its limiting form.
Thus the energy density away from the source on the boundary remains
finite  and well defined in the boundary condition limit even though
the total energy diverges.

\bigskip
\stepcounter{subsection}
\leftline{\bf 3.4 Gau{\ss}ian Background Field}
\bigskip

\begin{figure}
\centerline{
\epsfig{file=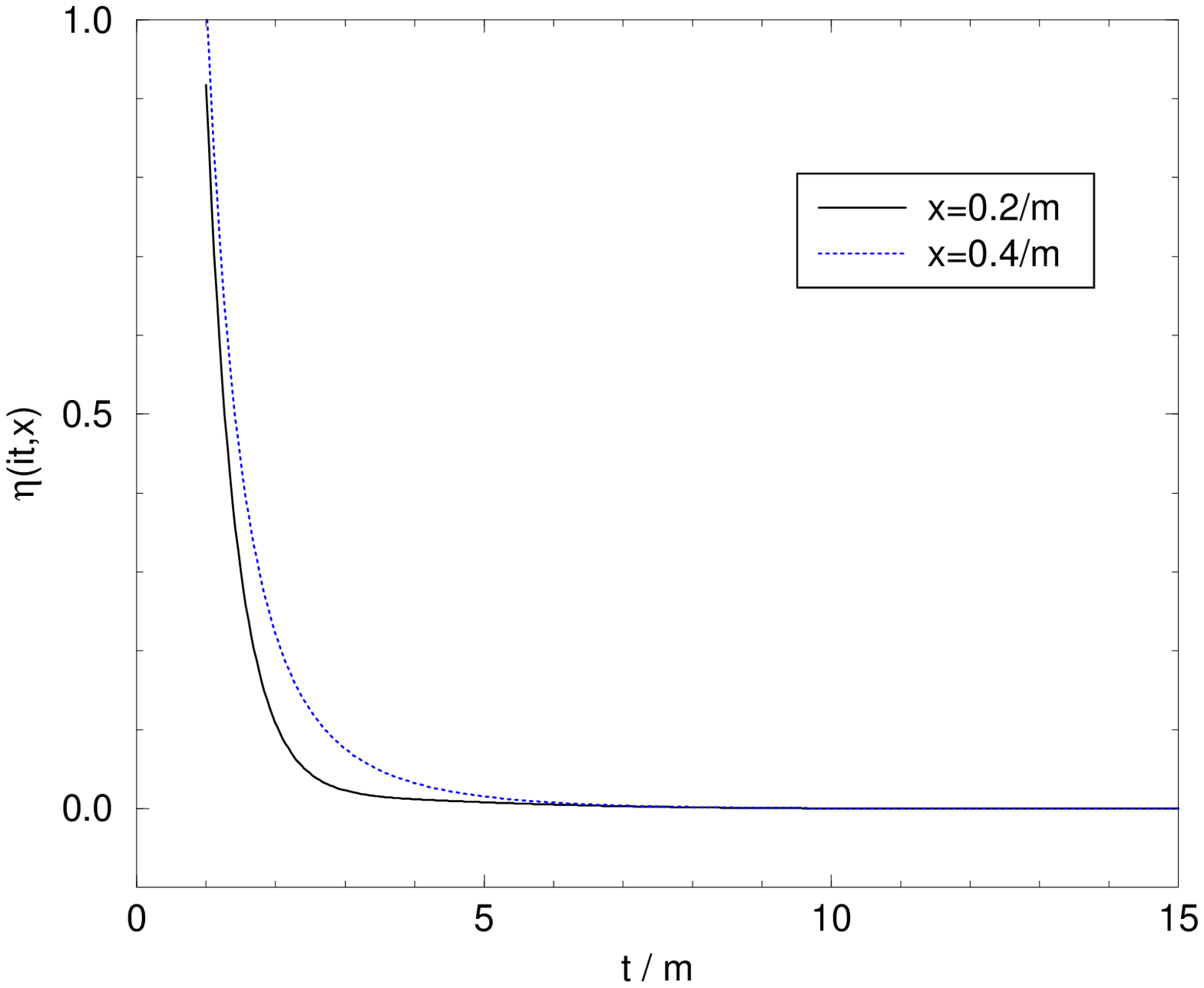,height=5.0cm,width=6cm}
\hspace{2cm}
\epsfig{file=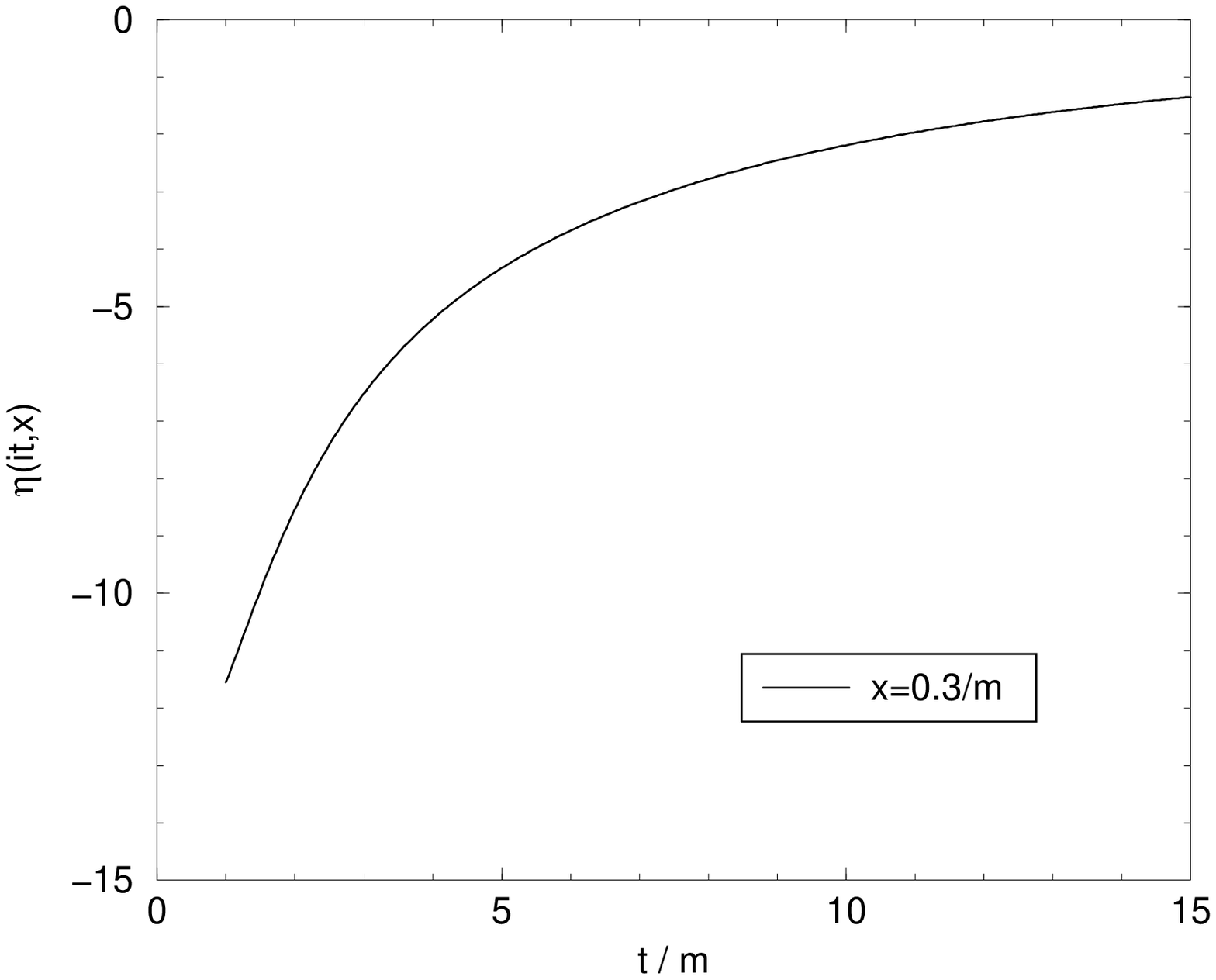,height=5.0cm,width=6cm}
}
\caption{\small The Born subtracted weight function $\eta(it,x)$ for the
Gau{\ss}ian background field, eq.~(\protect\ref{2dgauss}), as function
of the imaginary momentum $t$ at several positions $x$.  The
parameters are $a=0.3/m$, $w=0.03/m$ and $\lambda=3m$.}
\label{fig_new2}
\end{figure}

In this subsection we demonstrate how to calculate the energy density
associated with a Gau{\ss}ian background using our method.  This
problem is a warmup for the next Section, where we will use a tall,
narrow Gau{\ss}ian background as an approximation to the
$\delta$-function in higher dimensions, where the energy of a delta
function diverges even at finite coupling.  We choose a Gau{\ss}ian
background that approximates the $\delta$-function potential from the
previous subsection,
\begin{equation}
\sigma_{\rm G}(x)=\frac{A}{2}\left[e^{-\frac{(x-a)^2}{2w^2}}
+e^{-\frac{(x+a)^2}{2w^2}}\right]\,,
\label{2dgauss}
\end{equation}
in the limit $w\to 0$ where $A$ is chosen so that
the area under the Gau{\ss}ian is fixed
to $2\lambda$.  We substitute this background field
into the differential equations~(\ref{ODE1}), (\ref{ODE2})
and~(\ref{deqhP}) and obtain the wavefunctions $g$, $h_-$ and
$h_+$ numerically.
\begin{figure}
\centerline{
\epsfig{file=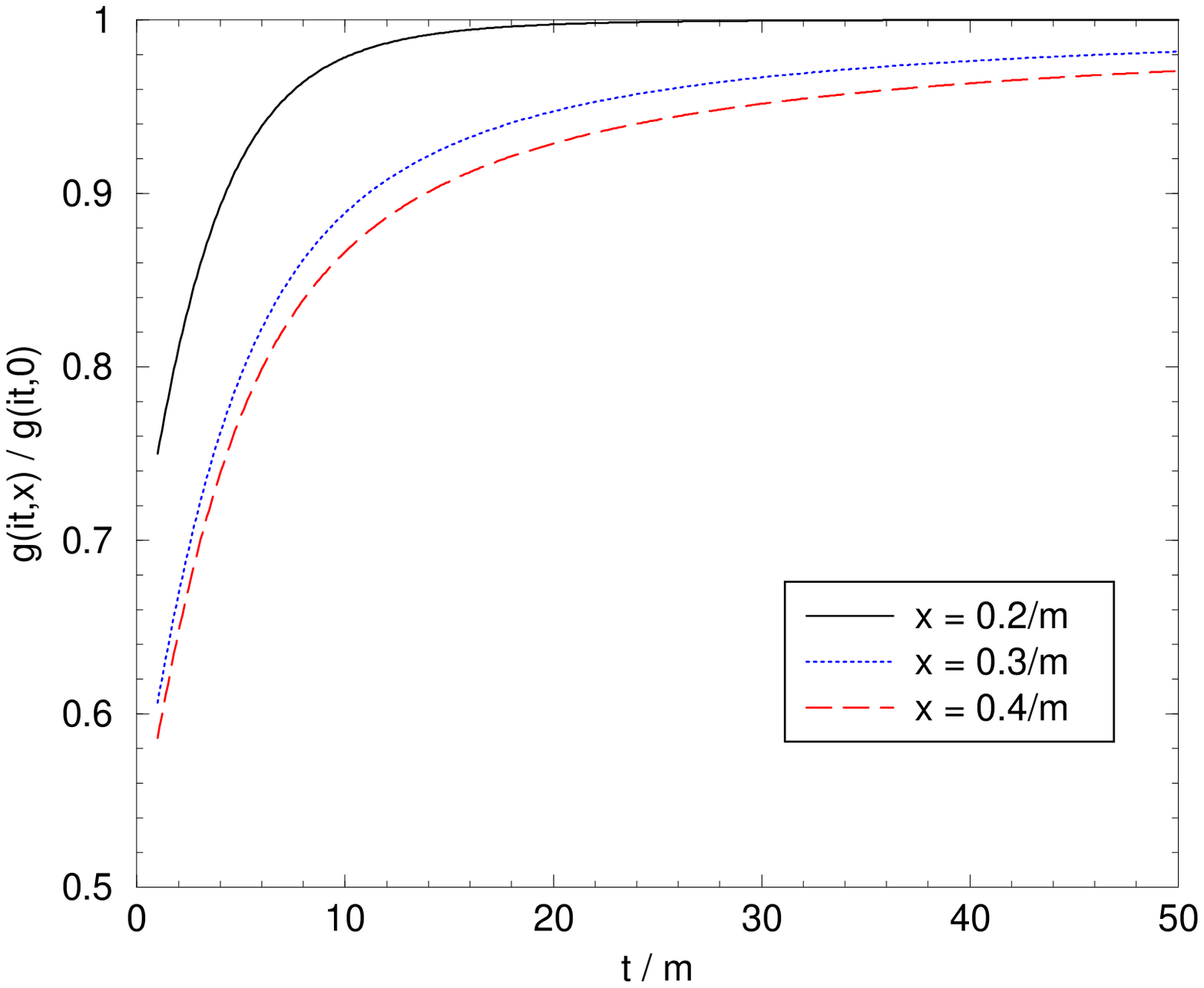,height=5.0cm,width=6cm}
\hspace{2cm}
\epsfig{file=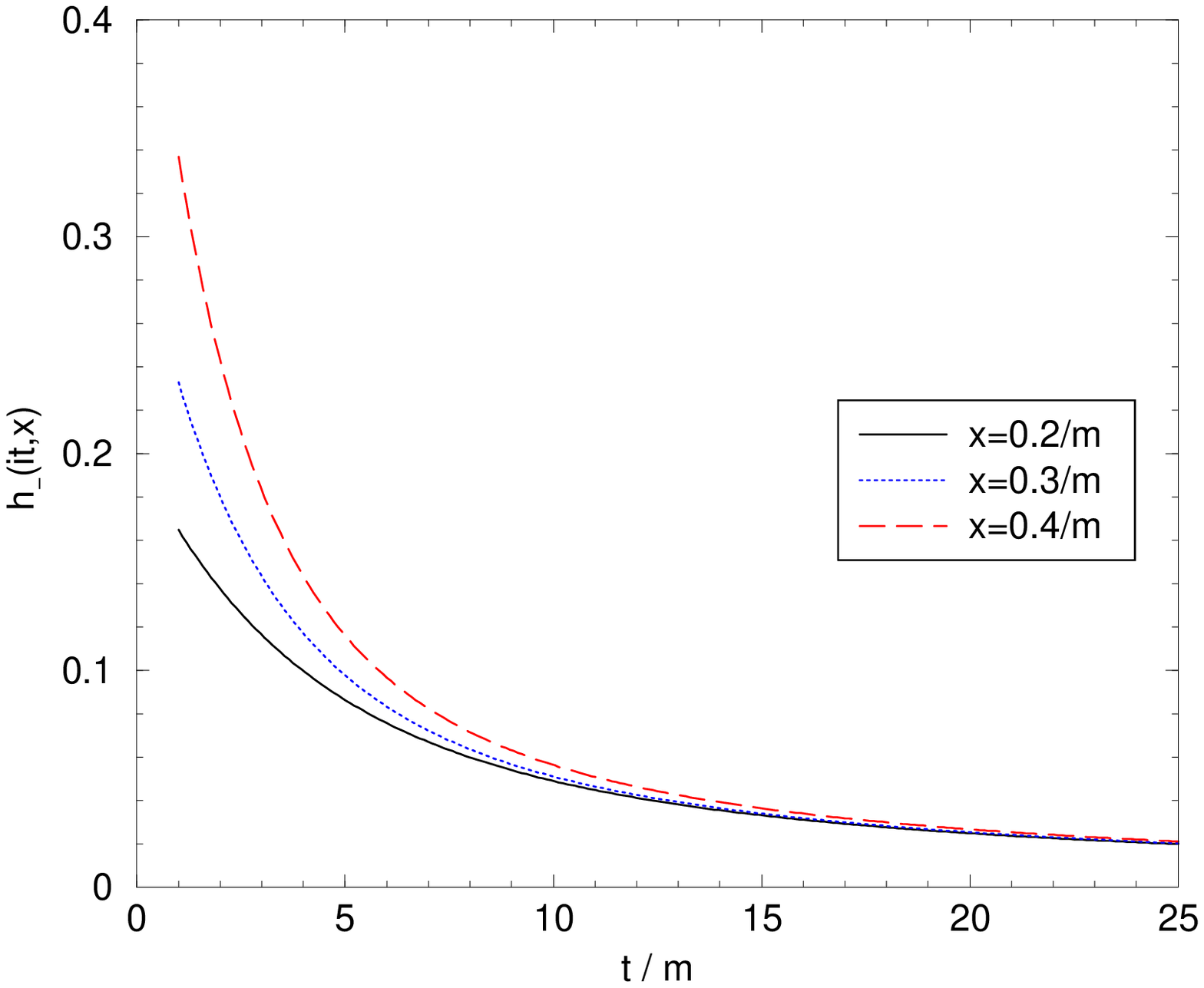,height=5.0cm,width=6cm}
}
\caption{\small Plots of $g(it,x)/g(it,0)$ and $h_-(it,x)$ as functions of
the imaginary momentum $t$ at several positions $x$.  The parameters
are $a=0.3/m$, $w=0.03/m$ and $\lambda=3m$.}
\label{fig_new1}
\end{figure}
As promised, these functions are well--behaved on the positive
imaginary $k$ axis. At large momenta $t$, we expect $h_{-}(it,x)$ to
approach the free solution, which decays asymptotically as $(1+
e^{-t|x|})/2t$. This behavior can be seen clearly in
Fig~\ref{fig_new1}. By the same argument, we expect the ratio
$g(it,x)/g(it,0)$ to approach one at large $t$, which is confirmed in
Fig~\ref{fig_new1}. The functions required for the symmetric channel
behave similarly.  The Born subtracted continuum contribution to the
energy density can be written in the form of eq.~(\ref{ebs}), where
now $\eta(it,x)$ is computed numerically from $g(it,x)$ and
$h_{\pm}(it,x)$.  In Fig.~\ref{fig_new2} we plot $\eta(it,x)$ for the
Gau{\ss}ian background for several values of $x$. 
Since $\sigma$ is nonzero everywhere, $\eta$ falls like
$1/t^3$ at large $t$ for all $x$.  We see that the contribution for
$|x\pm a|\gg w$, where the potential $\sigma$ is very small, is much
smaller than the contribution from $x\approx a$, where the potential
is peaked.

\begin{figure}
\centerline{
\epsfig{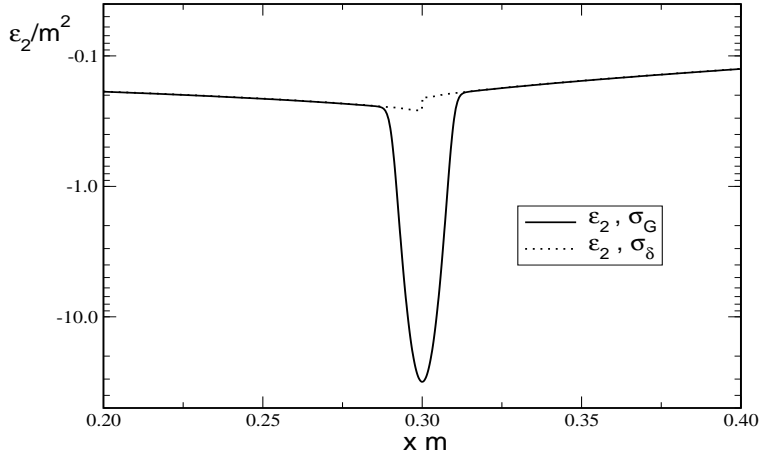}
}
\caption{\small Comparison of the energy densities for a
narrow Gau{\ss}ian, $\sigma_G$ with $w=0.003/m$, and a
$\delta$-function background, $\sigma_\delta$, located at $a=0.3/m$.
The coupling is $\lambda=3m$. Note that we have omitted the $\delta(x-a)$
piece for $\sigma_\delta$.  Its coefficient is such that when it is
added to the dotted line the areas under both curves are equal.}
\label{fig_gauss}
\end{figure}

The energy density we obtain for a tall, narrow Gau{\ss}ian is shown
in Fig.~\ref{fig_gauss}, together with the result from the previous
subsection for the $\delta$-function background.  The latter case also
has a $\delta$-function contribution to the energy, which is not
shown.  For $|x-a| \gapprox 3w$ the energy densities for these two background
fields are indistinguishable.  For  $|x-a| \lapprox 3w$ the energy
density for the Gau{\ss}ian background becomes dominated by a strongly peaked
contribution proportional to $\sigma(x)$ (approximating the
$\delta$-function contribution).

Finally, as a check on our numerical calculation, we have
verified that the total energy computed directly from
eq.~(\ref{evac1d}) agrees with the spatial integral of the energy
density.

\bigskip
\stepcounter{section}
\leftline{\large\bf 4. An Example in Two Space Dimensions}
\bigskip

Our numerical methods become particularly valuable as we move to
higher dimensions.  Our renormalization procedure and numerical methods
allow us to study a sequence of sources and examine the behavior of
the energy and energy density as one attempts to take the boundary
condition limit.  In two dimensions we still only need the
renormalization counterterm  proportional to $\sigma(\vec x)$.  To go
on to three dimensions would require an additional Born subtraction and a
counterterm proportional to $\sigma(\vec x)^{2}$.

We begin by discussing some peculiarities of the differential
equations~(\ref{ODE1}) and~(\ref{ODE2}) for the auxiliary radial
functions $g_\ell$ and $h_\ell$ necessary to handle the $s$-wave in
two dimensions.  Then we describe the computation of the total energy
and energy density for a Gau{\ss}ian background of width $w$ centered
on a circle of radius $a$, which illustrates our general conclusions
about the nature of the Casimir problem.

\bigskip
\stepcounter{subsection}
\leftline{\bf 4.1 Boundary Conditions for Radial Functions}
\bigskip

In two space dimensions the angular momentum $\ell$ can be restricted
to $\ell\ge0$ with degeneracy $N_{\ell}=2$ for $\ell\neq 0$ and $N_\ell=1$ for
$\ell=0$.\footnote{In the formula for $N_\ell$, given after
eq.~(\ref{ebare}), we first have to take $\ell=0$ and {\it
subsequently} $n\to2$ for the proper treatment of the singularity
in $\Gamma(n+\ell-2)$ for this case~\cite{FGJW_Levi}.}  In each
channel with $\ell > 0$, we can solve the radial problem by following
the steps described in Section 2.2. Specifically, $\nu \equiv \ell - 1
+ \frac{n}{2} = \ell$ and
\begin{equation}
\xi_\ell(z) = - \frac{d}{dz} \ln \left(\sqrt{z}\,K_\ell(z)\right)
= \frac{\ell-\frac{1}{2}}{z} + \frac{K_{\ell-1}(z)}{K_\ell(z)}\quad
\qquad (\ell \ge 0)\,,
\end{equation}
where $K_{-1} \equiv K_1$.  These functions are smooth for positive
arguments $z>0$ and quickly approach unity for $z \gg 1$.  For
$\ell>0$ they have a simple pole at $z=0$, $\xi_\ell(z)\simeq (\ell -
\frac{1}{2})/z$.  However, the functions $g_\ell$ and $h_\ell$ are
regular near $r=0$,
\begin{equation}
g_\ell(it,r) = g_\ell(it,0) + \mathcal{O}(r^2)
\qquad\quad
h_\ell(it,r) = r +\mathcal{O}(r^3)
\label{gh_approx}
\end{equation}
canceling the potential singularities at $r=0$ in eqs.~(\ref{ODE1})
and~(\ref{ODE2}) coming from $\xi_\ell(z)$.

For $\ell = 0$ it appears from eqs.~(\ref{factorw})
and~(\ref{Green3}) that both $h_0$ and the Green's function $G_0$
become ill--defined for $\nu = \ell = 0$. Ultimately, the source of
the problem can be traced to the subleading logarithmic term in the
Bessel functions,
\begin{equation}
w_0(kr) = i \sqrt{\frac{\pi}{2}kr} H_0^{(1)}(kr)
\quad {\rm and}\quad
\xi_0(z) = - \frac{1}{2z} - \frac{1}{z \ln z} + \mathcal{O}(z)\,.
\end{equation}
The subleading term shows up as a singularity in $\xi_0'(z)$, while it
is harmless in $\xi_0(z)$ itself. The solution is to integrate the
differential equation~(\ref{ODE1}) for $g_0$ from $r=\infty$ down to a
point arbitrarily close to $r=0$ and then get $g_0(it,0)$ from
eq.~(\ref{gh_approx}).

For the regular solution, we simply omit the normalization $1/2\nu$
in the {\it ansatz} (\ref{factorw}) and set
\begin{equation}
\phi_0(k,r) = \frac{\sqrt{k} h_0(k,r)}{w_0(k r)}\, .
\end{equation}
Then eq.~(\ref{Green3}) becomes
\begin{equation}
G_0(r,r,k) = \frac{g_0(k,r) h_0(k,r)} {g_0(k,0)}\,.
\end{equation}
The resulting differential equation for $h_0$ is still
eq.~(\ref{ODE2}), but the boundary conditions at $r=0$ must be
modified: It is straightforward to verify that
\begin{equation}
h_0(it,r) = - r\ln t r
\qquad{\rm and}\qquad
h_0'(it,r) = - \left[1+\ln tr\right]
\quad{\rm as}\quad r\to0
\label{h0limit}
\end{equation}
removes\footnote{The logarithmic behavior is actually expected from
the free solution: $rI_0(tr)K_0(tr)\to\ln 2 - \gamma - r\ln(tr)$
as $r\to0$, {\it cf.} eq.~(\ref{expODE}).} all small $r$
singularities in eq.~(\ref{ODE2}).  We use eq.~(\ref{h0limit}) to
integrate the differential equation~(\ref{ODE2}) numerically starting
at an arbitrarily small but nonzero value $r$.

It is then also straightforward, although tedious, to expand
$g_\ell$ and $h_\ell$ in powers of the background field $\sigma$
according to eq.~(\ref{expODE}). We thus obtain the Born subtracted
local spectral density, eq.~(\ref{state_dens}), for each angular
momentum quantum number~$\ell$.

The final computational ingredient in eq.~(\ref{fundamental}) is the
renormalized contribution from the Feynman diagrams.  In two space
dimensions one Born subtraction suffices.  Then we have to add
back in the two Feynman diagrams in Fig.~\ref{diagrams}.  In the
no--tadpole renormalization scheme the local diagram,
eq.~(\ref{direct}), is exactly canceled by the counterterm and we only
need to add back the two--point contribution, eq.~(\ref{final}).
The corresponding Feynman parameter integral is readily
computed to be
\begin{equation}
\epsilon_{\rm FD}(r) = \frac{m^2 r}{16 \pi}\int_0^\infty dp
J_0(pr) \tilde{\sigma}(p) \left[ p + \frac{p^2 - 4 m^2}{2 m}
\arctan\frac{p}{2m}\right]\, ,
\end{equation}
where
\begin{equation}
\tilde{\sigma}(p) = 2 \pi \int_0^\infty dr \,r J_0(p r) \sigma(r).
\label{dia}
\end{equation}

\bigskip
\stepcounter{subsection}
\leftline{\bf 4.2 Numerical Results for a Gau{\ss}ian
Background Field}
\bigskip

We now apply this method and discuss the numerical results for the
total energy and energy density of a Gau{\ss}ian background
field in two dimensions. We consider
\begin{equation}
\sigma(r) = \lambda A e^{-\frac{(r-a)^2}{2 w^2}}\, .
\label{gauss_back}
\end{equation}
This profile describes a ring of radius $a$ and thickness $w$.  The
normalization $A$ is chosen such that $\int_0^\infty \sigma(r) dr =
\lambda,$ giving $\sigma(r) \to \lambda\delta(r-a)$ as $w\to
0$. Sending the coupling strength $\lambda$ to infinity subsequently
imposes the Dirichlet boundary condition on a circle of radius $a$.

\begin{figure}
\centerline{ \epsfig{file=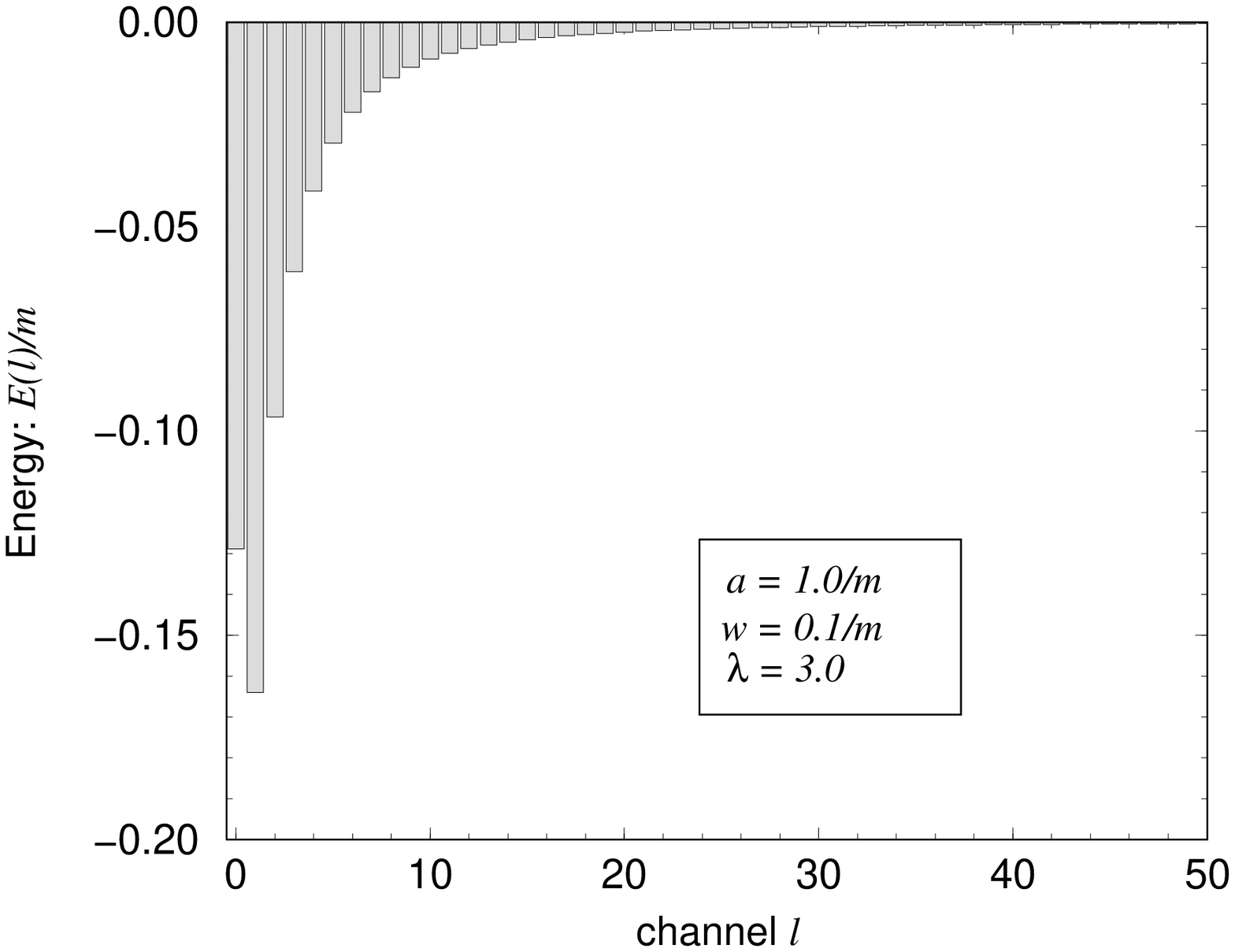,height=6.0cm,width=7cm}
\hspace{1cm} \epsfig{file=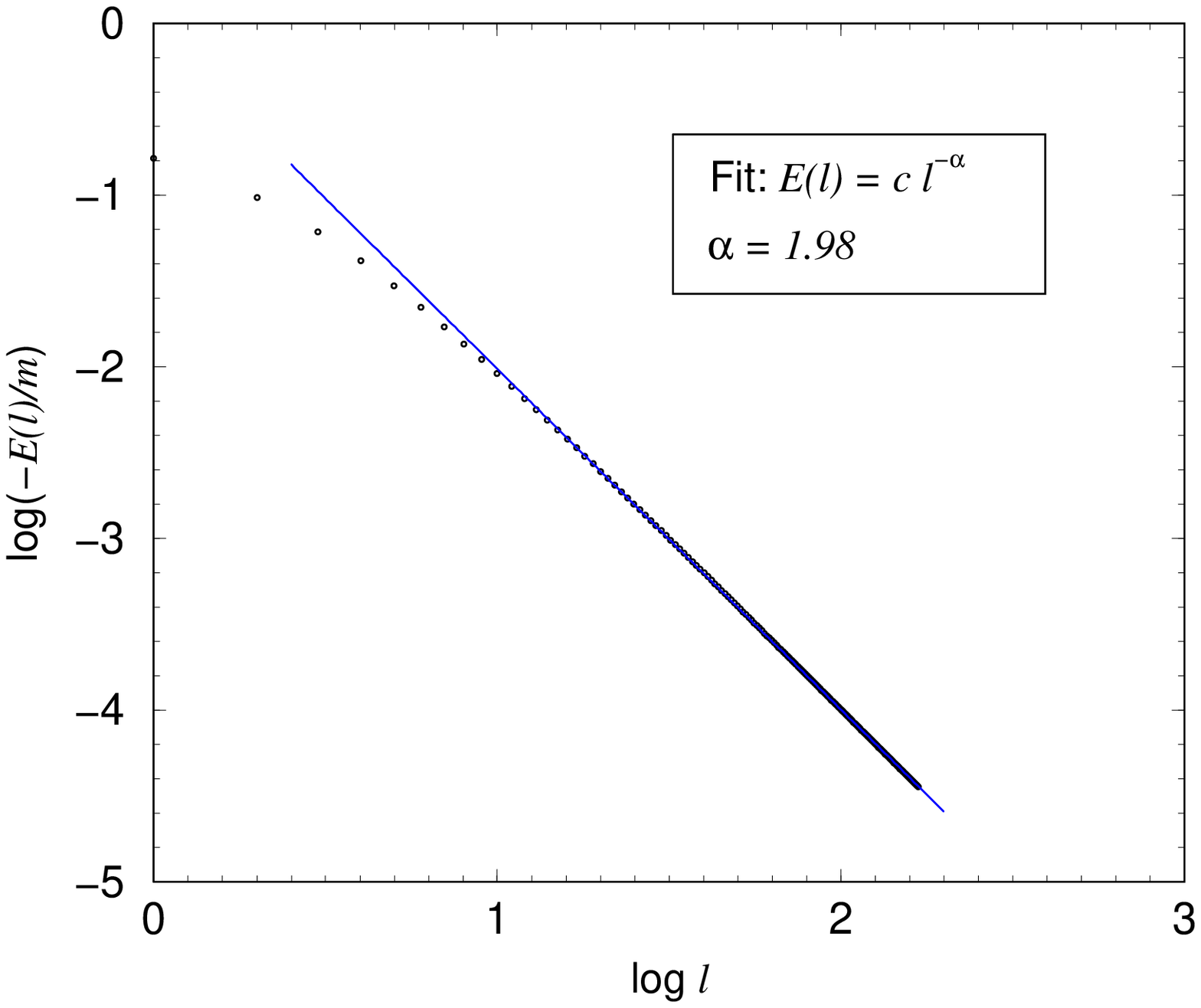,height=6.0cm,width=7cm} }
\caption{\small Contributions $E_{\ell}(\lambda,a,w)$ to the total
energy from various angular momentum channels $\ell$.  The logarithmic
plot on the right is consistent with the asymptotic behavior
$E_{\ell}\sim 1/\ell^2$.  The parameters are $a=1.0/m$, $w=0.1/m$ and
$\lambda=3m$.}
\label{total1}
\end{figure}

We first discuss the total energy $E[\sigma]=E(\lambda,a,w)$ computed
from eq.~(\ref{evac3}).  Note that no Feynman diagram contribution has
to be added to the total energy because in the no-tadpole scheme $E_{\rm
FD}^{(1)}+E_{\rm CT}=0$.  Fig.~\ref{total1} shows the total energy
contribution $E_{\ell}(\lambda,a,w)$ of the angular momentum channel
$\ell$, for a Gau{\ss}ian ring of radius $a=1.0/m$ and width
$w=0.1/m$.  Note that the $\ell=0$ channel is suppressed relative to
$\ell > 0$ by the degeneracy factor $N_0/N_\ell = \frac{1}{2}$.
Unlike the case in three dimensions, however, higher angular
momentum channels are \emph{not} favored by a further increasing
degeneracy, so the contributions decay rapidly with increasing $\ell
\ge 1$.  The largest contribution always comes from $\ell=1$.
The logarithmic plot in Fig.~\ref{total1} confirms that the asymptotic
region is reached quickly and that the decay of the individual
contributions is consistent with $E_{\ell} \propto 1/\ell^2$ for $\ell
\ge 5$.  While this clearly shows the finiteness of the renormalized
quantum energy (for fixed $a$ and $w$), the convergence is too slow to
allow us to sum the $\ell$-series directly.  Instead, we use a speedup
technique related to Richardson's approach \cite{NR}, which is
particularly powerful in the present case, where the asymptotic
behavior sets in quickly.  We have checked that the results of this
method are very robust and independent of the details of the
extrapolation technique used.  A sufficiently high precision in the
energy and density calculation can then be achieved by using only
approximately 30 channels.  For details on this summation technique,
see Appendix~C.

The computation of the total energy can be sped up even further by
subtracting the second Born approximation in eq.~(\ref{fundamental})
and adding back the corresponding second--order Feynman diagram. With
this (over--)subtraction, the large $\ell$ behavior of
$E_{\ell}(\lambda,a,w)$ changes to $1/\ell^{3}$.  The relevant Feynman
diagram is easily computed,
\begin{equation}
E_{\rm FD}^{(2)}[\sigma] = \frac{1}{32 \pi^2} \int_0^\infty dp
\tilde{\sigma}(p)^2 \arctan\frac{p}{2m}
\label{order2}
\end{equation}
where $\tilde{\sigma}$ denotes the Fourier transform, eq.~(\ref{dia}).
We have verified that this method gives the same result as before.

\begin{figure}
\centerline{
\epsfig{file=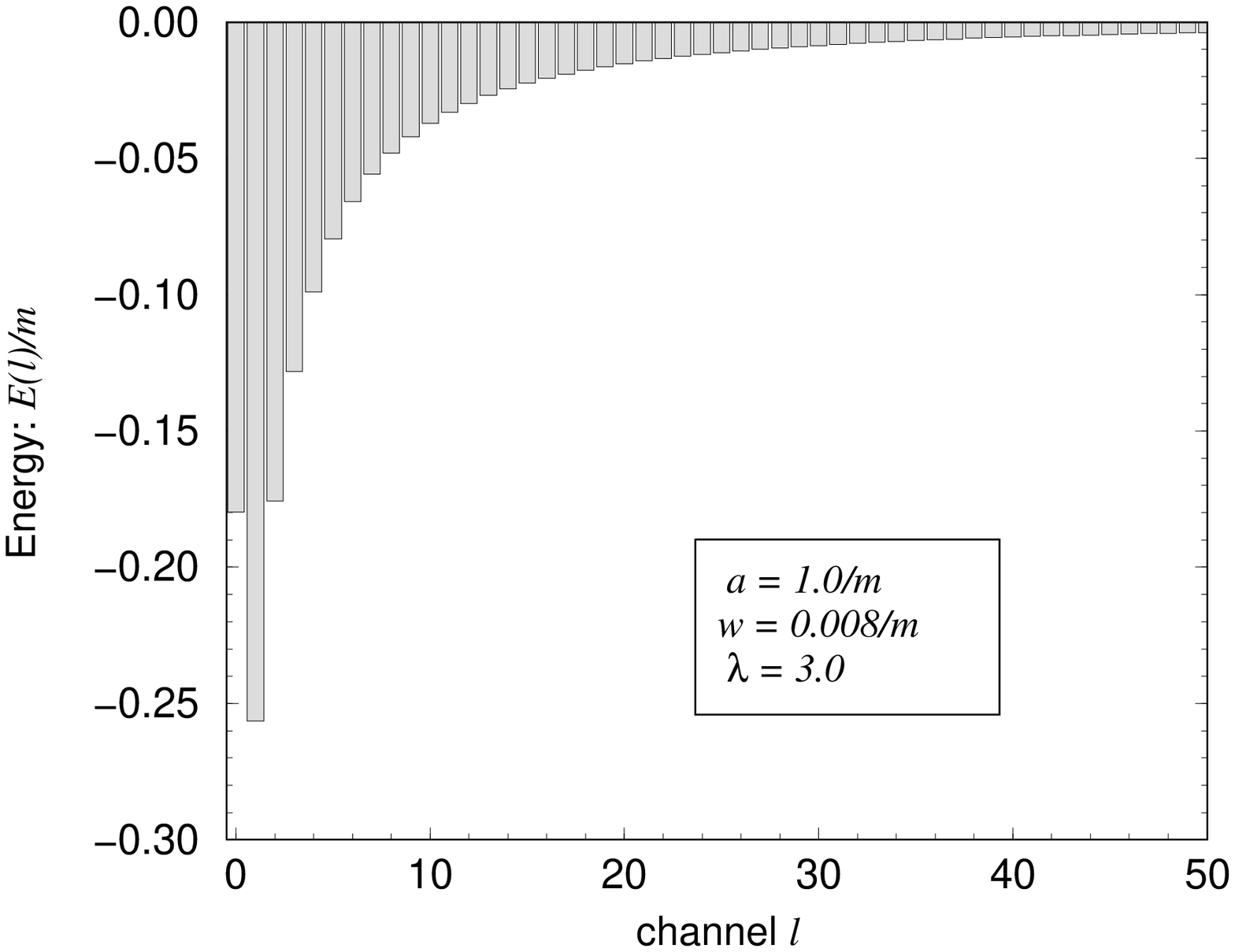,height=6.0cm,width=7cm}
\hspace{1cm}
\epsfig{file=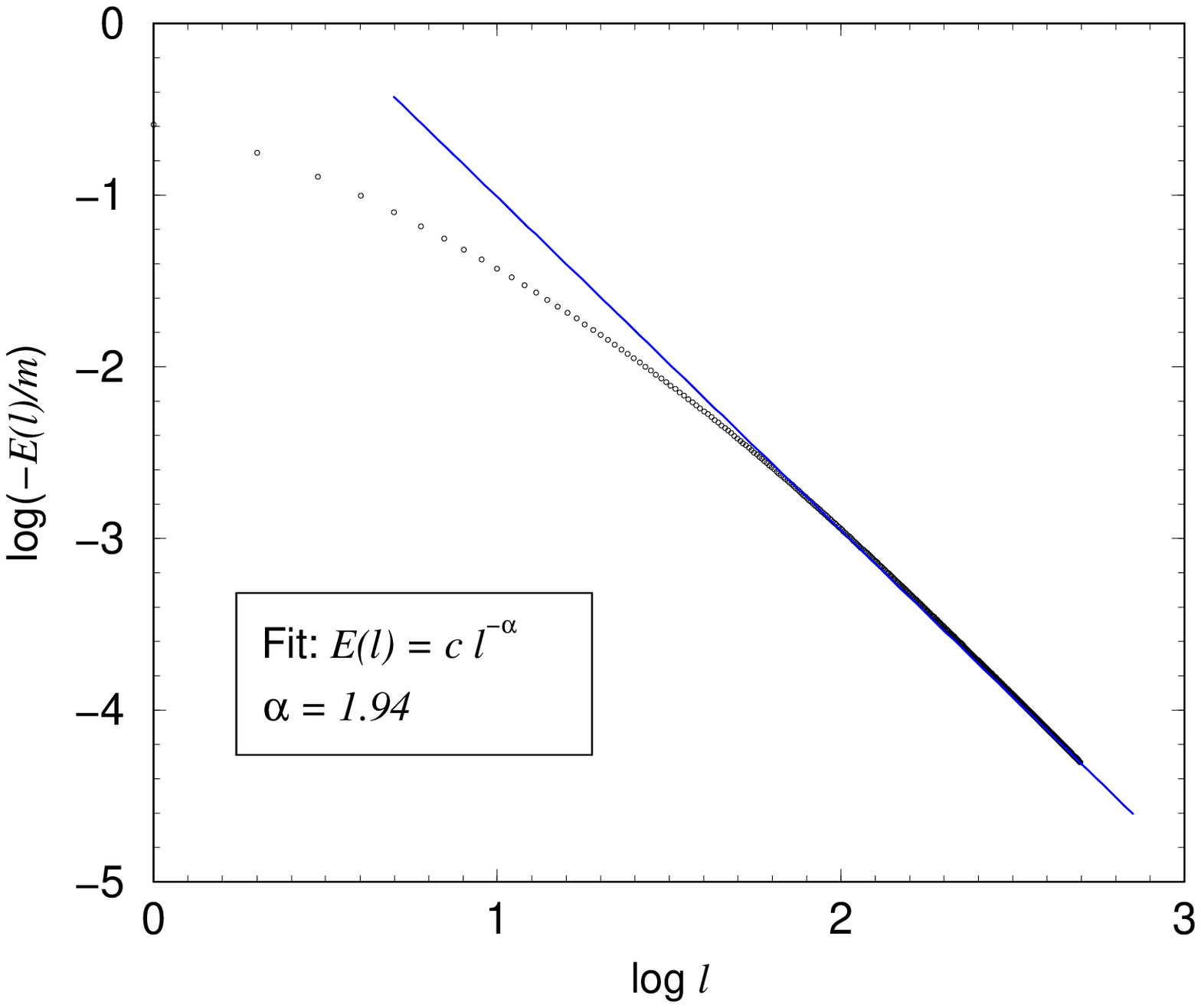,height=6.0cm,width=7cm}
}
\caption{\small Contributions $E_{\ell}$ to the total energy from
various angular momentum channels $\ell$. The Gau{\ss}ian background is
much narrower than in Fig.~\ref{total1} and the asymptotic decay
$E_\ell\sim 1/\ell^2$ is reached for much larger channels $\ell$
only.  The parameters are $a=1.0/m$, $w=0.008/m$ and $\lambda=3m$.}
\label{total2}
\end{figure}

As the background field approaches a $\delta$-function we find by
explicit numerical calculation that the total Casimir energy
diverges. To illustrate the problem we consider a very narrow
configuration with $w=0.008/m$. As can be seen by comparing
Figs.~\ref{total2} and \ref{total1}, $E_{\ell}$ decreases more slowly
with $\ell$ when $w$ is smaller.  For any finite width $w$ the energy
contributions $E_\ell$ eventually decrease steeply enough to yield a
convergent sum and a finite total energy, but this asymptotic regime
is only reached for much larger $\ell$ when the width is small. This
is a typical non--uniform behavior:  For any fixed width, there is
always a channel $\ell_0$ large enough such that the energies $E_\ell$
decay as $1/\ell^2$ for $\ell \ge \ell_0$. On the other hand, for
every fixed channel $\ell$ we can always find a width $w$ small enough
such that we are far from the asymptotic region. Eventually, as $w\to
0$ the asymptotic region is never reached and the total energy diverges.

\begin{figure}
\centerline{
\epsfig{file=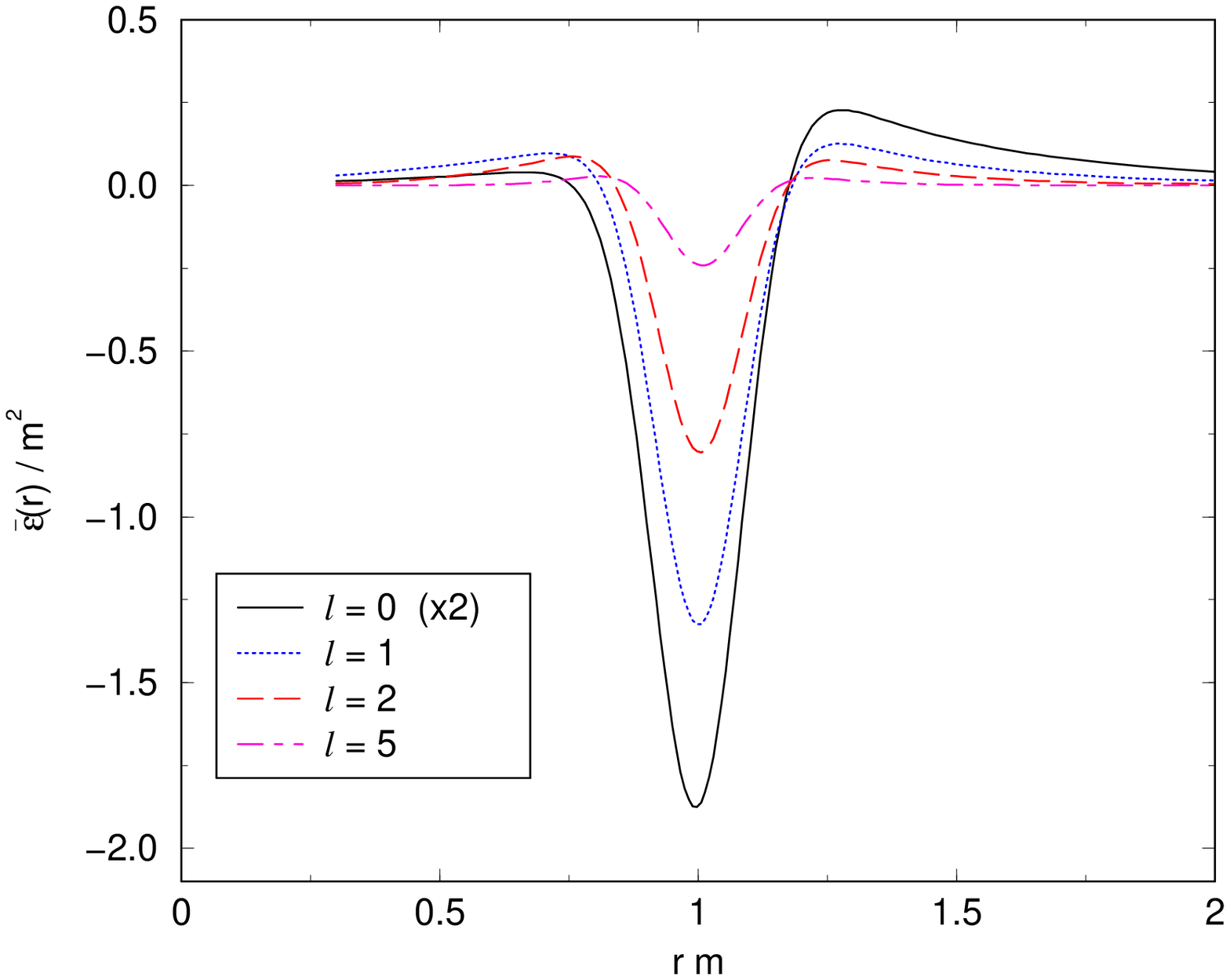,height=6.0cm,width=7cm}
\hspace{1cm}
\epsfig{file=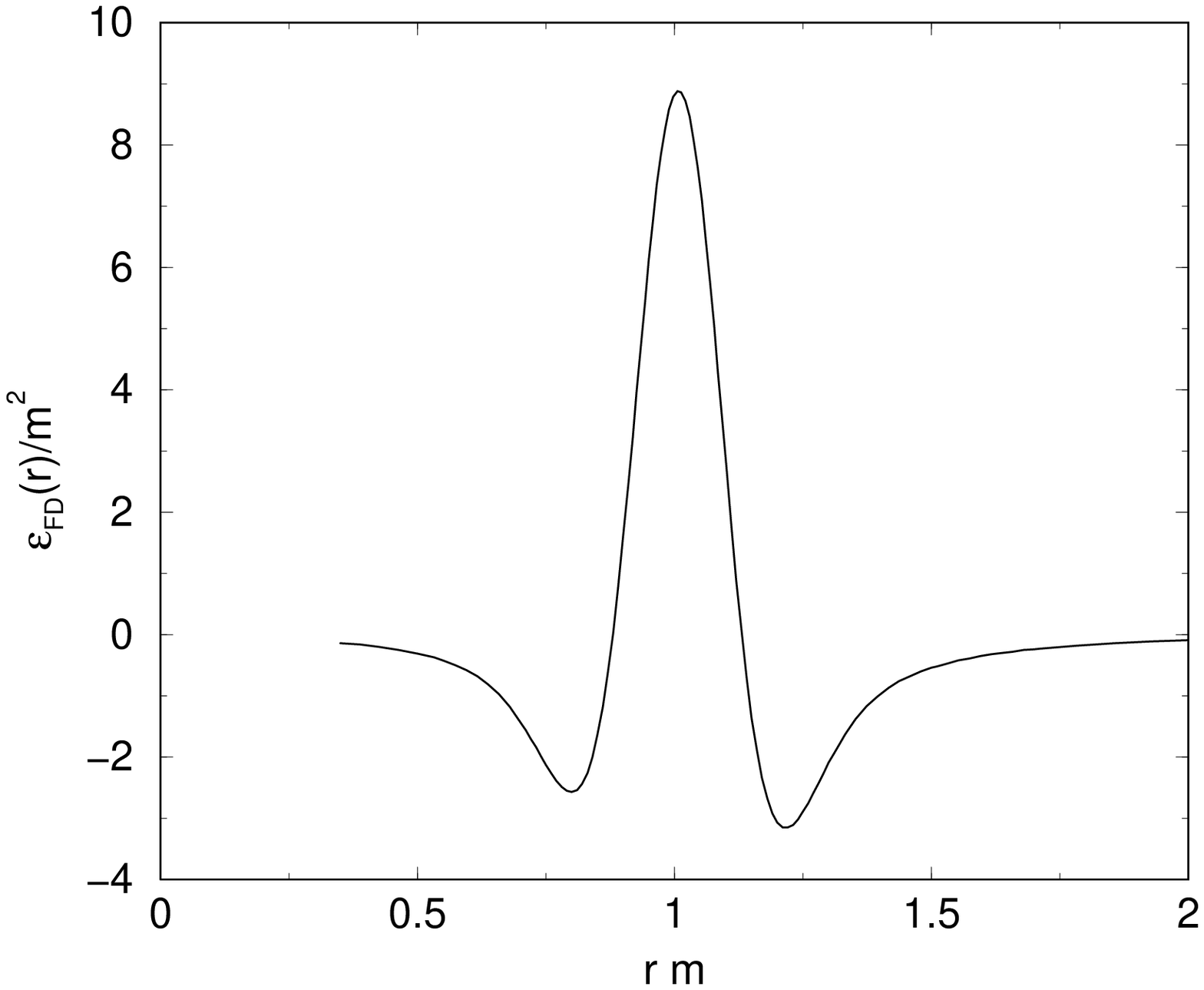,height=6.0cm,width=7cm}
}
\caption{\small Left panel: Contributions of various
angular momentum channels to the Born subtracted
energy density~$\bar{\epsilon}(r)$. Note that we have multiplied
the $\ell=0$ contribution by a factor of 2 to better display the
secular decrease with $\ell$.  Right panel: The Feynman diagram
contribution $\epsilon_{\rm FD}(r)$. The parameters for both panels
are $a=1.0/m$, $w=0.1/m$ and $\lambda=3m$. See also
Fig.~\ref{fig_3dwidth} for the total energy density $\epsilon(r)$
found by combining the components displayed here.}
\label{final_dens}
\end{figure}

One can also understand this divergence in terms of the second Born
approximation, which we separated out in the oversubtraction procedure
above.  For the singular background, $\sigma(r) = \lambda\delta(r-a)$,
this diagram contributes
\begin{equation}
E_{\rm FD}^{(2)}[\lambda\delta(r-a)] = \frac{\lambda^2
a^2}{8}\int_0^\infty dp J_0^2(a p) \arctan\frac{p}{2m}
\label{e2}
\end{equation}
which diverges logarithmically at large $p$ just as the sum over channels did
without oversubtraction. Oversubtraction shifts the divergence from
the $\ell$ sum into the momentum integral in the Feynman
diagrams.  Whichever choice we make, the divergence cannot be
renormalized away, despite the claims of \cite{Milton}; it is a
physical effect.  Its implications for the usual Casimir calculations
will be studied elsewhere~\cite{letter}.

Next we examine the radial energy density $\epsilon(r)$.  
As a numerical check, we have verified that its integral over
$r$ agrees with the total energy computed directly from
eq.~(\ref{evac3}).  In Fig.~\ref{final_dens} we display the
contributions of various angular momentum channels to the Born
subtracted energy density, $\bar{\epsilon}(r)$, for a relatively wide
profile, $w=0.1/a$.  Generically the contributions to the energy
density from various angular momentum channels vary in magnitude but
have similar radial shapes.  As for the total energy, the $\ell=0$
channel is suppressed by a degeneracy factor $1/2$; the largest
contribution comes from $\ell=1$ with all higher channels decaying
quickly with increasing $\ell$.  For narrower widths the decay with
$\ell$ is slower, as we expect from the preceding discussion of the
total energy. The smooth asymptotic behavior allows us again to use an
extrapolation method in order to speed up the sum over channels.

\begin{figure}
\centerline{
\epsfig{file=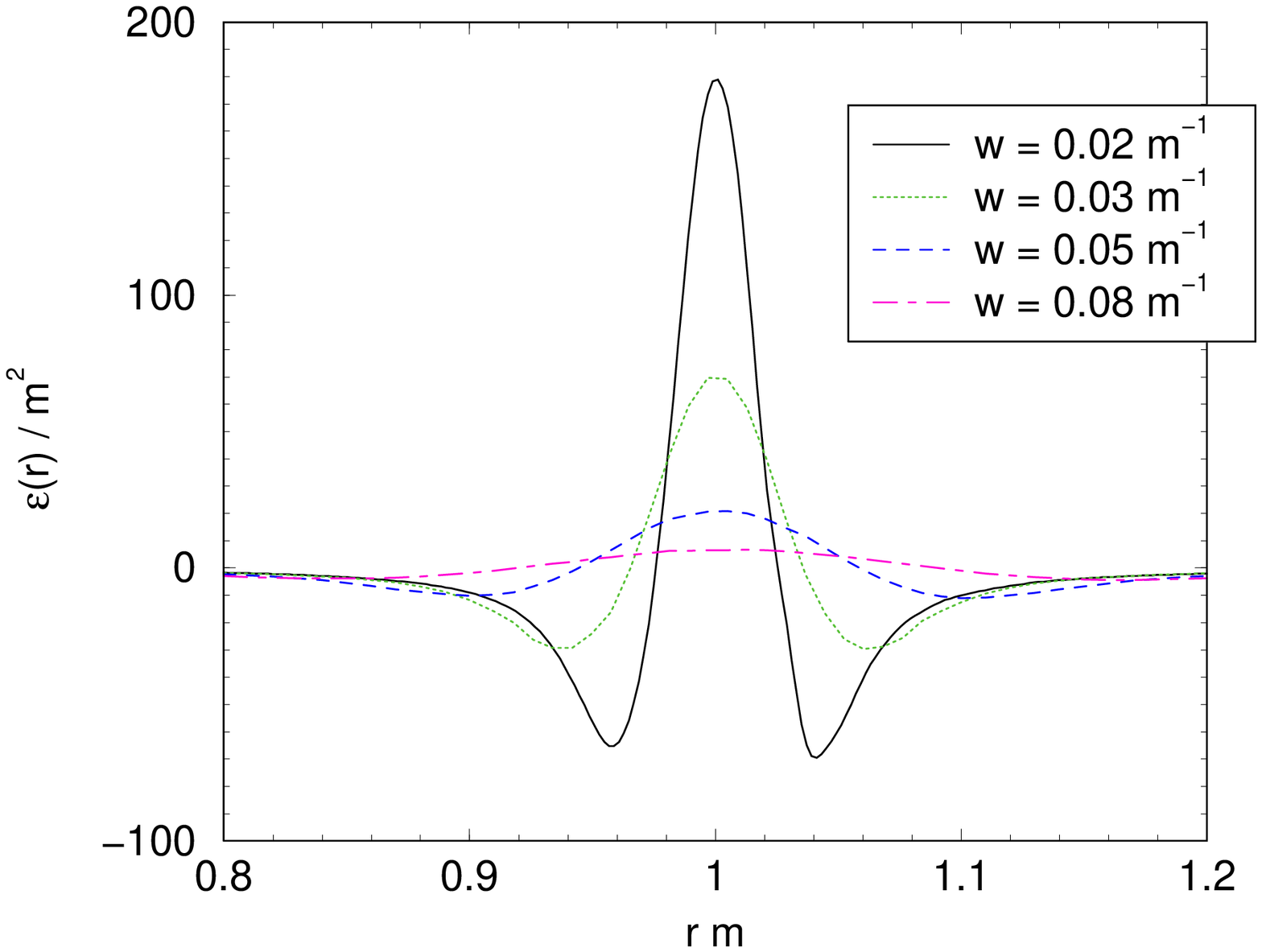,height=6.0cm,width=7cm}
\hspace{1cm}
\epsfig{file=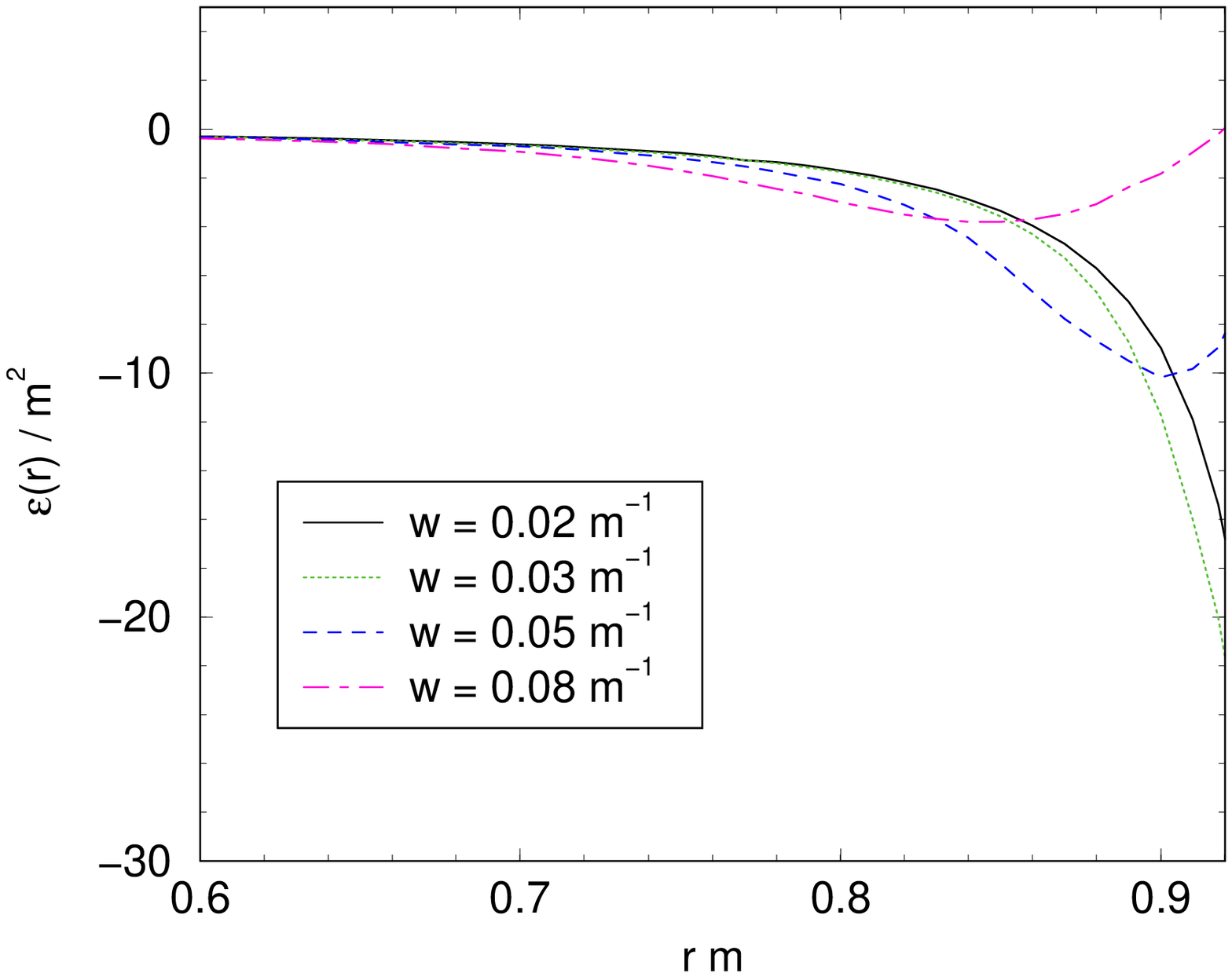,height=6.0cm,width=7cm}
}
\caption{\small The energy density in units of $m^2$ for the Gau{\ss}ian
background field located at $a=0.1/m$ with strength $\lambda=3.0m$
for various values of the width $w$. The left panel shows the energy
density for all radii. The right panel focuses on the region where the
densities can be seen to converge toward the limiting form.}
\label{fig_3dwidth}
\end{figure}

Figure \ref{final_dens} also shows the renormalized Feynman diagram
contribution in the right panel.  Since the integral over all
radii $r$ is zero, the peak at the location of the background profile
must be compensated by negative densities on both sides.  The sign and
width of the peak essentially reproduce the Gau{\ss}ian background
itself, while the similar peak in the density $\bar{\epsilon}(r)$ has
the opposite sign and reduces the overall magnitude of the density. As
can be seen from the total energy density $\epsilon(r)$ in
Fig.~\ref{fig_3dwidth}, combining all the contributions still yields a
pronounced peak at the location of the background. As we decrease the
width of $\sigma(r)$, the peak increases in magnitude and becomes
narrower.  In the $\delta$-function limit it develops a non--integrable
singularity, unlike the case in one dimension.

\begin{figure}
\centerline{
\epsfig{file=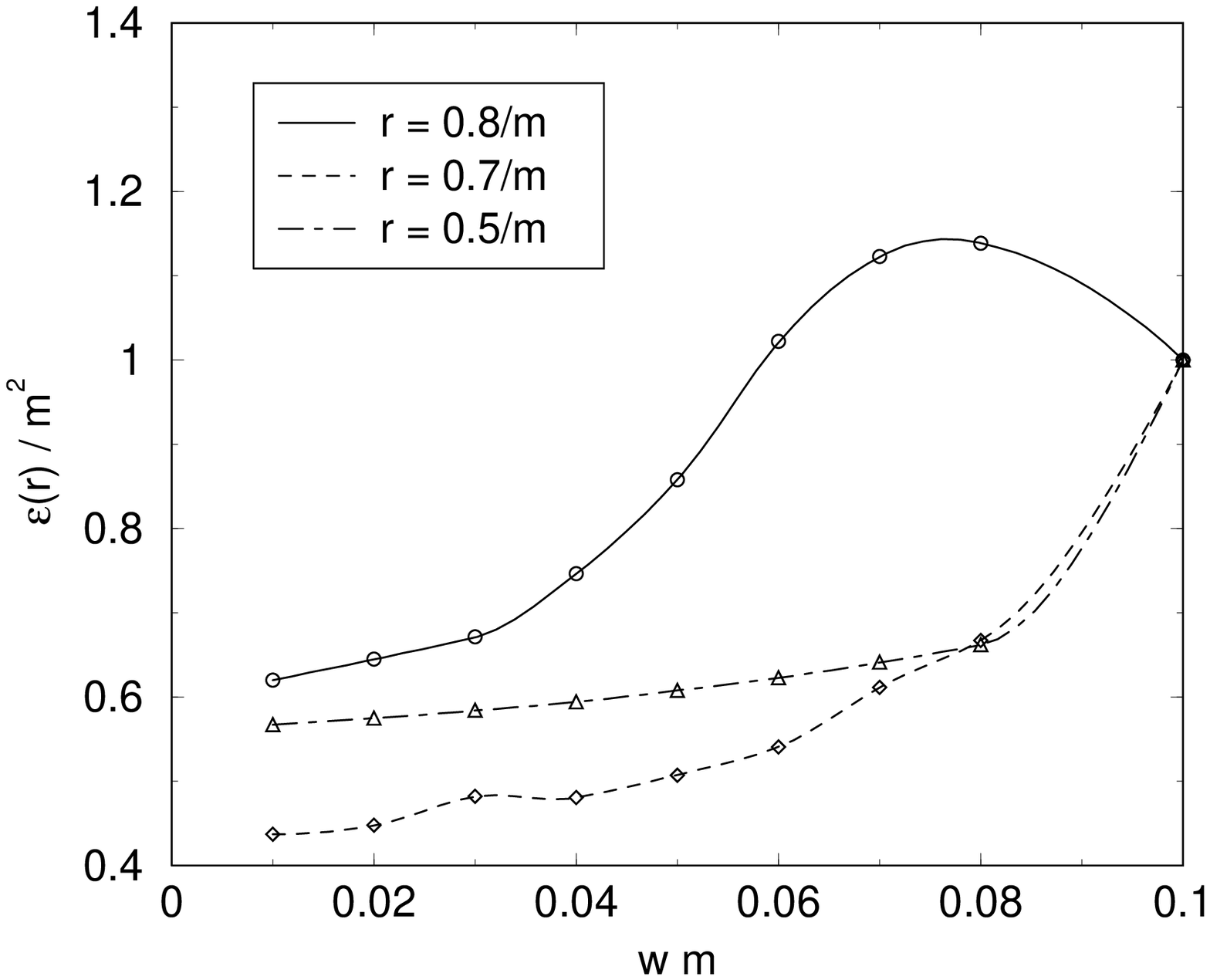,height=6.0cm,width=7cm}
\hspace{1cm}
\epsfig{file=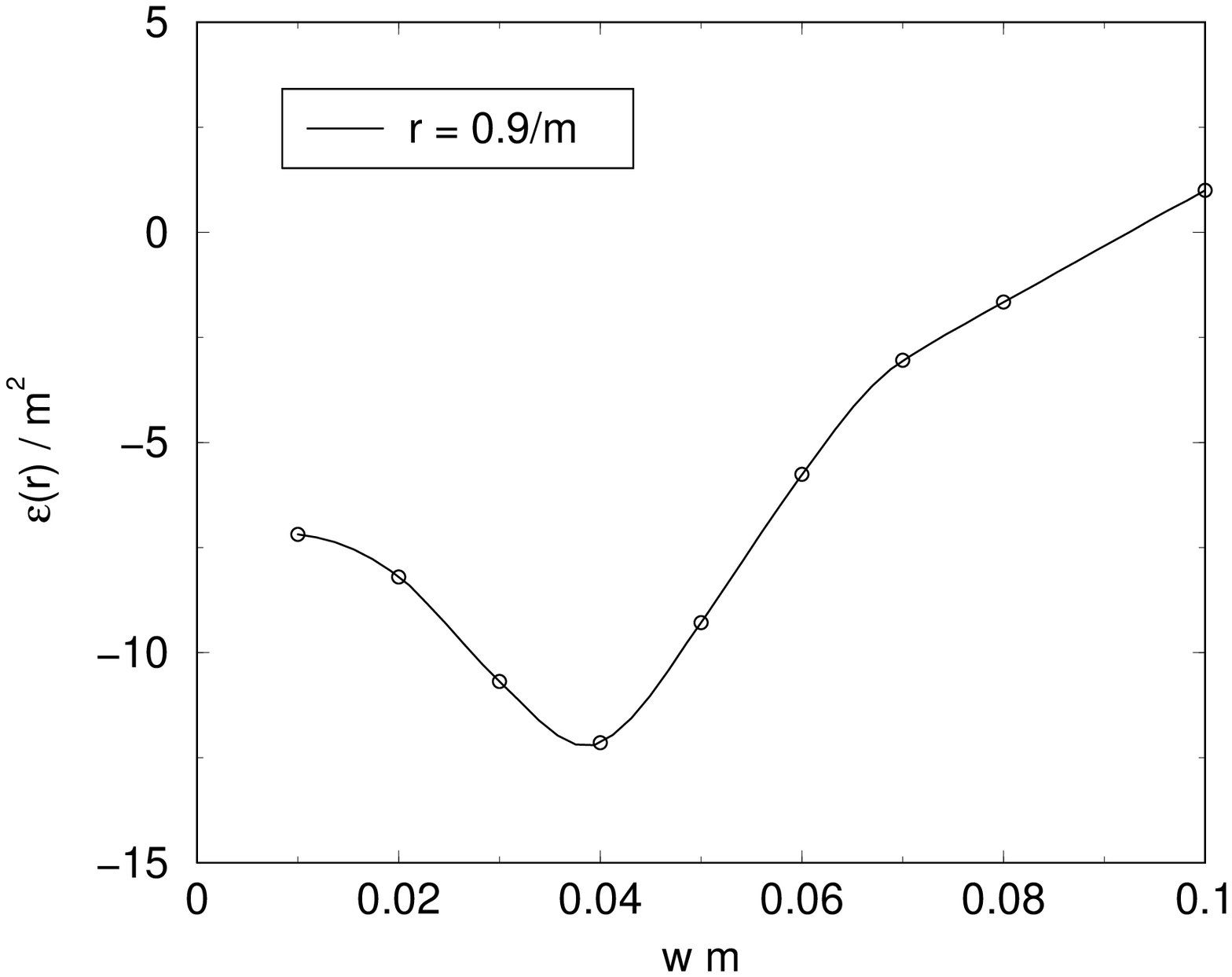,height=6.0cm,width=7cm}
}
\caption{\small The energy density
of the Gau{\ss}ian background field located at $a=1.0/m$ and
with strength $\lambda=3.0m$. The figure shows the energy
densities at various values of $r$ as functions of the width,
$w$. The densities are scaled to $\epsilon(r)$ at $w=0.1/m$.}
\label{fig_3dlimit}
\end{figure}

Figs.~\ref{fig_3dwidth} and \ref{fig_3dlimit} display the dependence
of the energy density for various values of $r$ on the width of the
Gau{\ss}ian background field.  We see the non--uniform approach to the
limit $w\to 0$.  In particular, it is clear that the energy density
approaches a finite limit at any fixed $r$ as $w\to 0$, even though the
limiting function diverges as $r\to a$.
Sufficiently far away from the location of the Gau{\ss}ian,
around $|r-a| \gapprox 3w$, the energy density quickly approaches the
limiting form, corresponding to that of a $\delta$-function
background.  In contrast, at $r=a$, no such limiting value exits.
This behavior can also be seen in Fig.~\ref{fig_3dlimit}, where we
display the energy density at various values of $r$ as function of the
width. The limiting form is approached more rapidly the farther one is
away from the Gau{\ss}ian peak.

The Dirichlet boundary condition is achieved only by taking
$\sigma(r)\to \lambda\delta(r-a)$, followed by the limit
$\lambda\to\infty$.  However, since the renormalized Casimir energy
diverges as $\sigma$ approaches a $\delta$-function, it is impossible to
define the Casimir energy for any sharp background field, much less
one with infinite coupling. Thus the Dirichlet--Casimir energy of a
circle in two dimensions does not exist.  But all is not lost:  the
energy density away from the source approaches a finite limit as the
background becomes sharp.

\bigskip
\stepcounter{section}
\leftline{\large\bf 5. Summary and Outlook}
\bigskip

In this paper we have developed an efficient method to
compute one--loop energies and energy densities in renormalizable
quantum field theories.  Starting from a Green's function formalism,
we employ Born subtractions to implement definite renormalization
conditions, and express the result in a form that allows for direct,
efficient calculation after analytic continuation to the imaginary $k$
axis.  This method allows us to study the limit in which smooth background
fields approach idealized boundary conditions, illuminating
subtleties of the process by which standard Casimir calculations are
realized in renormalized quantum field theory.  We find that although
the Casimir force between rigid bodies and energy densities away from
the boundaries are well--defined observables, quantities like the
Casimir energy have divergences that cannot consistently be
renormalized.  These divergences have physical consequences in
examples such as the stress on a circle, where we have to compare two
configurations for which the divergent contributions differ.  In these
cases, the physical effects depend strongly on the details of the source.

We have also shown rigorously that the total energy, as obtained from the 
integrated energy density, can be expressed as the
integral over single particle energies weighted by the change in the
density of states, which is given by the derivative of the phase shifts.

There are many possible applications of our method. It can be
generalized to fermions~\cite{Sundberg} and to situations where the
energy spectrum is asymmetric about zero.   In the latter case we would
have to add the different contributions from the two sides of the cut
in Fig.~\ref{contourfig}.  In such systems the background field can
carry nonzero charge density, since the contribution to the charge
density from positive and negative energy modes will no longer cancel.
It could also be calculated with our approach.  We could consider the
electromagnetic field in the presence of conductors by decomposing it
into a scalar field obeying Neumann boundary conditions plus a scalar
field obeying Dirichlet boundary conditions.  We have seen in this
paper how to implement Dirichlet boundary conditions with a background
potential, and the Neumann boundary condition can be treated in a similar
way~\cite{Farhi}.  Therefore we are equipped to reconsider problems
such as the surface tension of a conducting sphere in QED
\cite{Boyer}.  Our calculational method is well suited to soliton
calculations, generalizing \cite{SUNYSB} beyond the case of
reflectionless potentials.  Finally, in cases like the Gross--Neveu or
Nambu--Jona--Lasinio models, where one introduces the background
potential as a non--dynamical auxiliary field coupled to a dynamical
field, one can search for solitons as self--consistent solutions to the
auxiliary field equations of motion \cite{Al96,Ba98}.  These equations
of motion are obtained by first integrating out the dynamical field,
leaving an effective action for the auxiliary field.  This procedure
generates expressions involving the full Green's function similar to
those we have considered here, which could be computed efficiently
with our methods.

\bigskip
\leftline{\large \bf Acknowledgments}
\bigskip

\noindent
\enlargethispage{4mm}
We gratefully acknowledge discussions with E.~Farhi and
K.~D.~Olum.  N.~G. and R.~L.~J. are supported in part by the
U.S.~Department of Energy (D.O.E.) under cooperative research
agreements~\#DE-FG03-91ER40662 and~\#DF-FC02-94ER40818.  M.~Q. and
H.~W. are supported by the Deutsche Forschungsgemeinschaft under
contracts~Qu 137/1-1 and~We~1254/3-2.
\vskip1cm

\appendix
\bigskip
\stepcounter{chapter}
\leftline{\large\bf Appendix A: Density of States and the Jost Function}
\bigskip 

In this Appendix, we demonstrate the relation (\ref{app1}) between the
integral over $r$ of the local spectral density and the Jost function
\begin{equation}
2 \int_0^\infty dr\,\left[\rho_\ell(k,r)\right]_0 = 
\frac{2 k}{i} \int_0^\infty dr\,\left[G_\ell(r,r,k) - G^{(0)}_\ell(r,r,k)
\right] = i \frac{d}{dk} \ln F_\ell(k)\,,
\label{appa_1}
\end{equation}
which is valid everywhere in the upper half--plane $\mathsf{Im}\,k >
0$.  This formula can be used both to establish an efficient
computation of the total energy on the imaginary axis, and also to
relate the density of states for real $k$ to the phase shift.
 
We start by differentiating the Wronskian of the Jost solution,
$f_\ell(k,r)$, and the regular solution, $\phi_\ell(k^\prime,r)$
\cite{Newton},
\begin{equation}
\frac{d}{dr} W\left[f_\ell(k,r),\phi_\ell(k^\prime,r)\right]
=\left(k^2-k^{\prime2}\right)f_\ell(k,r)\phi_\ell(k^\prime,r)
\label{eqna1}
\end{equation}
where all quantities in this relation are analytic for 
$\mathsf{Im}\,k>0$.  We integrate both sides from $r=0$ to $r = R$.
Since the regular solution $\phi_\ell(k,r)$ becomes $k$--independent
at small $r$, we can compute the boundary term at $r=0$ by replacing
$k'$ with $k$ and using the standard Wronskian,
\begin{equation}
W[f_\ell(k,r),\phi_\ell(k,r)] = \left(-k\right)^{\frac{1}{2}-\nu} F_\ell(k)\,,
\end{equation}
giving
\begin{equation} 
W\left[f_\ell(k,R),\phi_\ell(k^\prime,R)\right] =
\left(-k\right)^{\frac{1}{2}-\nu} F_\ell(k)+
\left(k^2-k^{\prime2}\right)
\int_0^R dr\, f_\ell(k,r)\phi_\ell(k^\prime,r)\,.
\label{eqna2}
\end{equation}
Next we differentiate with respect to $k$, set $k^\prime=k$, and use
the representation (\ref{Green2}) for the Green's function to obtain
\begin{equation}
\frac{\left(-k\right)^{\nu-\frac{1}{2}}}{F_\ell(k)}
W\left[\dot{f}_\ell(k,R),\phi_\ell(k,R)\right]=
\frac{\nu-\frac{1}{2}}{k}+\frac{\dot{F}_\ell(k)}{F_\ell(k)}
+2k\int_0^R dr\, G_\ell(r,r,k)\, ,
\label{eqna3}
\end{equation}
where $\dot{f}_\ell(k,R)\equiv\frac{d}{dk}f_\ell(k,R)$
and $\dot{F}_\ell(k)\equiv\frac{d}{dk}F_\ell(k)$.
To eliminate the first term on the right--hand side, we subtract the
same equation for the non--interacting case, giving
\begin{eqnarray}
&&\frac{\left(-k\right)^{\nu-\frac{1}{2}}}{F_\ell(k)}
W\left[\dot{f}_\ell(k,R),\phi_\ell(k,R)\right]-
\left(-k\right)^{\nu-\frac{1}{2}}
W\left[\dot{f}^{(0)}_\ell(k,R),\phi^{(0)}_\ell(k,R)\right]&
\cr
&&\hspace{3cm} = \frac{\dot{F}_\ell(k)}{F_\ell(k)} 
+2k\int_0^R dr\, \left[G_\ell(r,r,k)
-G^{(0)}_\ell(r,r,k)\right]\,.
\label{eqna4}
\end{eqnarray}
To complete the proof, we have to show that the left hand side of 
eq.~(\ref{eqna4}) vanishes as $R\to\infty$.  To see this, we write the
boundary condition (\ref{jost_bc}) for the Jost solution in the form
\begin{equation}
f_\ell(k,R) = w_\ell(kR) \,\left[1 + \mathcal{O}(R^{-1})\right]\,,
\qquad\quad R \to\infty
\end{equation}
which can also be inferred from the integral equation obeyed by $f_\ell(k,r)$.
Differentiating with respect to $k$ and using the asymptotics of 
the free Jost solution $w_\ell(kR)$, 
\[
\dot{w}_\ell(kR) = \frac{d}{dk}\,w_\ell(k R) = i R\,\,w_\ell(k R)
\,\left[ 1 + \mathcal{O}(R^{-2})\right]\,, 
\]
it is easy to show the asymptotic behavior
\begin{equation}
\dot{f}_\ell(k,R) = i R\, f_\ell(k,R)\,\left[1 + 
\mathcal{O}(R^{-2})\right]\,.
\end{equation}
The first term on the left--hand side of eq.~(\ref{eqna4}) can thus be
estimated by
\begin{eqnarray}
\lefteqn{\frac{(-k)^{\nu-\frac{1}{2}}}{F_\ell(k)}\,W[\dot{f}_\ell(k,R),
\phi_\ell(k,R)] =} \nonumber \\
&& \left\{i R \,\frac{(-k)^{\nu-\frac{1}{2}}}{F_\ell(k)}
W[f_\ell(k,R),\phi_\ell(k,R)]  - i\,\frac{(-k)^{\nu-\frac{1}{2}}}
{F_\ell(k)}\,f_\ell(k,R)\phi_\ell(k,r)\right\} \,\left[1 + 
\mathcal{O}(R^{-2})\right] \nonumber \\
&& = - i \left[ R + G_\ell(R,R,k) \right] \,\left[1 + 
\mathcal{O}(R^{-2})\right]\,,
\label{lead1}\end{eqnarray}
where we have used the Wronskian of $f_\ell$ and $\phi_\ell$ and 
the definition of the Green's function, eq.~(\ref{Green2}).
Subtracting the analogous equation in the free case, the term
proportional to $R$ drops out and we are left with
\begin{eqnarray}
&&\frac{\left(-k\right)^{\nu-\frac{1}{2}}}{F_\ell(k)}
W\left[\dot{f}_\ell(k,R),\phi_\ell(k,R)\right]-
\left(-k\right)^{\nu-\frac{1}{2}}
W\left[\dot{f}^{(0)}_\ell(k,R),\phi^{(0)}_\ell(k,R)\right]\cr
&&\hspace{3cm} = -i \left[ G_\ell(R,R,k) - G_\ell^{(0)}(R,R,k)\right] 
\,\left[1 + \mathcal{O}(R^{-1})\right]\,.
\end{eqnarray}
We estimate the large--$R$ behavior of the difference
$\Delta_\ell(k,R) \equiv G_\ell(R,R,k) - G_\ell^{(0)}(R,R,k)$
from eqs.~(\ref{lead1}) and (\ref{eqna4}),
\begin{equation}
- i\,\Delta_\ell(k,R) \, [1 + \mathcal{O}(R^{-1})] = 
\frac{\dot{F}_\ell(k)}{F_\ell(k)} + 2 k \int^R_0dr\,\Delta_\ell(k,r)\,.
\label{klose}
\end{equation}
It is then straightforward to solve for $\Delta_\ell(k,R) \simeq 
i \,[C_\ell(k) + \mathcal{O}(1)R]\,\exp(2 i k R)$ at large $R$, 
where $C_\ell(k)$ is an $R$--independent integration constant.
The left--hand side vanishes exponentially for all
$\mathsf{Im}\,k > 0$ as $R\to\infty$, so in this limit we obtain
eq.~(\ref{appa_1}), which completes the proof.

It is also instructive to explore the consequences of this result on
the real axis.  If we let $k$ approach the real axis from above, we see that
to leading order in $R^{-1}$, the integral equation (\ref{klose})
allows for a solution going like $R \exp[2 i k R]$,
which violates the bounds in eq.~(\ref{bounds}) if $k$ is
real.  Thus the coefficient of this term must actually be zero, leaving
\begin{equation} 
\frac{\dot{F}_\ell(k)}{F_\ell(k)} + 2k \int_0^R dr\,\left[G_\ell(r,r,k)
- G_\ell^{(0)}(r,r,k)\right] = C_\ell(k) \exp(2 i k R)
\left[1 + \mathcal{O}(R^{-1})\right]
\label{lead2}
\end{equation}
which oscillates as $R\to\infty$ when $k$ is real.  As is typical for
continuum problems, we must specify that the limit where $k$ becomes
real is taken after the integral over $r$ to eliminate the
contribution from these oscillations at the upper limit of integration.

On the real axis, we can relate the Jost function to the
phase shift by eq.~(\ref{beckenbauer}).  Taking the imaginary part of
eq.~(\ref{appa_1}) and using eq.~(\ref{dos}) yields the
relationship between the density of states and the phase shift,
\begin{equation}
\frac{1}{\pi}\frac{d\delta_\ell}{dk} = 
\frac{2k}{\pi} \mathsf{Im} \int_0^\infty \left(
G_\ell(r,r,k+i\epsilon)-G_\ell^{(0)}(r,r,k+i\epsilon)\right) dr =
\rho_\ell(k) - \rho_\ell^{(0)}(k) \,.
\label{realaxis_a}
\end{equation}
We can also rewrite eq.~(\ref{realaxis_a}) as
\begin{equation}
\frac{2}{\pi} \int_0^\infty dr \left(
\psi_\ell^*(k,r) \psi_\ell(k,r) - 
\psi_\ell^{(0)}{}^*(k,r) \psi_\ell^{(0)}(k,r)\right) =
\frac{1}{\pi}\frac{d\delta_\ell}{dk}
\label{normtophase}
\end{equation}
where the momentum on the left--hand side is understood to be defined
with the $i\epsilon$ prescription to eliminate the contribution from
the oscillations of the wavefunctions at spatial infinity.

\bigskip

\stepcounter{chapter}
\leftline{\large\bf Appendix B: Two Jost functions in One Dimension}
\bigskip

In this Appendix, we discuss a technique special to one dimension,
which applies even when the  background field  is not symmetric.
Scattering in one dimension can be described in terms of two Jost
solutions $f_\pm(k,x)$, which obey the boundary conditions
\begin{equation}
\lim_{x\to\pm\infty}f_\pm(k,x)e^{\mp ikx}=1\, .
\label{bc2J}
\end{equation}
The Green's function then is
\begin{equation}
G(x,y,k)=i\frac{T(k)}{2ik}f_+(k,x_>)f_-(k,x_<)\,,
\label{Gr2J}
\end{equation}
where $T(k)$ is the transmission coefficient.  For numerical
computations it is convenient to factor exponentials out of
$f_{\pm}(k,x)$.  We define
\begin{equation}
g_\pm(k,x)=e^{\mp ikx}f_\pm(k,x) \,,
\label{factor2J}
\end{equation}
which obey the differential equations
\begin{equation}
-g^{\prime\prime}_\pm(it,x) \pm 2tg^\prime_\pm(it,x) +
\sigma(x)g_\pm(it,x)=0
\label{deq2J}
\end{equation}
with the boundary conditions
\begin{equation}
\lim_{x\to\pm\infty} g_\pm(it,x)=1 \qquad {\rm and} \qquad
\lim_{x\to\pm\infty} g^\prime_\pm(it,x)=0\, ,
\label{bc2J1}
\end{equation}
and are bounded in the upper half $k$--plane.
The Green's function for equal arguments then reads
\begin{equation}
G(x,x,it)=\frac{T(it)}{2t}g_+(it,x)g_-(it,x)\, .
\label{Gr2J1}
\end{equation}
The transmission coefficient is most conveniently computed
using the Wronskian for $f_\pm$, yielding
\begin{equation}
G(x,x,it)=\frac{g_+(it,x)g_-(it,x)}
{g^\prime_+(it,0)g_-(it,0)-g^\prime_-(it,0)g_+(it,0)
+2tg_+(it,0)g_-(it,0)}\,,
\label{Gr2J2}
\end{equation}
for the Green's function at equal spatial arguments.  In this form, we
can compute the Green's function without encountering any oscillatory or
exponentially growing contributions.

\bigskip
\stepcounter{chapter}
\leftline{\large\bf Appendix C: Slowly Convergent Series}
\bigskip

In the examples studied in the main text, the sum over angular momentum
channels $\ell$ is slowly converging, typically going as
$1/\ell^2$ for large $\ell$.  From a numerical point of view, such a
decay is too slow for a direct summation as it requires to compute
$\mathcal{O}(1000)$ channels to get a relative precision of
$1\%$. Even worse, the narrow Gau{\ss}ian profiles relevant for
Casimir problems have a tendency to fall like $1/\ell$ before they
eventually turn over to a $1/\ell^2$ behavior at channels with large
angular momenta, $\ell \geq \ell_0 \gg 1$.

To sum such series efficiently using only a small number of channels,
we employ a technique related to Richardson's method \cite{NR}. Given
a slowly converging sum $\sum\limits_{\ell=0}^\infty a_\ell$, we
define a function $f(t)$ at the particular points $t_n$ by
\begin{equation}
f(t_n) \equiv \sum_{\ell=0}^{n} a_\ell\,,\qquad\qquad
t_n \equiv \frac{1}{n+1}\qquad(n=0,1,2,.\ldots)
\end{equation}
which gives the value of the full infinite series in the limit $t\to 0$.
We assume that $f(0)$ can be approximated by a smooth extrapolation
using the points $t_n$, which accumulate around $t=0$.  Thus given
values $a_0,\ldots,a_N$ from the first $N+1$ channels, we construct an
interpolating function $\tilde{f}$ of a fixed order $\nu$ through the last
$\nu+1$ points $t_{N},t_{N-1},\ldots, t_{N-\nu}$ (for the first few
channels $N < \nu$, so in those cases we take $\nu=N$). The
extrapolation to $t=0$ gives an approximation $\tilde{f}(0)$
for the sum of the series and an error estimate based on the assumption of a
smooth underlying function $f(t)$. While new channels are computed, we
monitor $\tilde{f}(0)$ and terminate if we have a satisfying error
estimate.  For our purposes, the best results are obtained from a
rational function (or Pad\'{e}) approximation, since the resulting
$\tilde{f}(t)$ has the tendency to anticipate the smooth behavior of
the underlying $f(t)$. To illustrate the method, let us look at the example
\begin{equation}
\sum_{\ell=1}^\infty \frac{\ell_0}{\ell (\ell + \ell_0)}
\label{sum_test}
\end{equation}
which imitates the behavior of our typical sums over channels: It starts
with $a_\ell \sim 1/\ell$ and turns into the converging $1/\ell^2$
behavior for $\ell  > \ell_0$. In the table below, we show the results
of our method with the standard parameters used in the main text. We
also give the number of channels actually summed and compare with the
results obtained by direct summation of the same number of channels.
\begin{table}[ht]
\caption{\small Results of our extrapolation method as
compared to direct summation for the test series eq.~(\ref{sum_test}).}
\medskip
\centerline{
\begin{tabular}{|c||c|c|c|c|c|c|c|}
\hline
$\ell_0$ & 5 & 10 & 50 & 500 & 1000 & 5000 & 10000 \\\hline\hline
exact & 2.2833 & 2.9290 & 4.4992 & 6.7928 & 7.4855 & 9.0945 & 9.7875 \\ \hline
interpol.& 2.2833 & 2.9290 & 4.4949 & 6.7385 & 7.3077 & 8.0560
& 8.2627 \\
\small{\textit{error [\%]}} & $2.2\cdot 10^{-6}$ & $2.38\cdot 10^{-5}$
& 0.095 &  0.80 & 2.37 & 11.4 & 15.6 \\\hline direct& 1.8941 & 2.2602
& 2.7483 & 3.9672 & 4.1653 & 4.2706 & 4.3684
\\
\small{\textit{error [\%]}} &  17.0 & 22.8 & 38.9 & 41.6 & 44.4 & 53.0& 55.3
\\ \hline
channels& 10 & 10 & 10 & 31 & 37& 40 & 44 \\ \hline
\end{tabular}}
\label{tab_sum}
\end{table}

As we can see from the table, our method improves the accuracy
dramatically. For small turning points $\ell_0$, the first $10$
channels are sufficient to find the exact value. As we go to higher
$\ell_0$, it becomes more difficult to predict the exact result from
the lowest terms which do not yet see the asymptotic decay. Even then,
with $\ell_0$ as large as $500$, we can still achieve a precision of
$<1\%$ using the first $\mathcal{O}(30)$ channels only.  The direct
summation is off by more than $40\%$ in this case. Eventually, for
very large $\ell_0 \simeq 10000$, the error of the Pad\'{e}
approximation increases up to $15\%$. This is not too surprising since
there is no trace of the eventual $1/\ell^2$ decay in the first $44$
channels summed. In order to predict the correct result within $1\%$
for the case $\ell_0 = 10000$, our method would in fact have to use
$\mathcal{O}(400)$ channels (but the direct summation would
still be off by $>30\%$ in this case).

\end{document}